\documentclass[a4paper,fleqn,usenatbib]{mnras}

\usepackage{newtxtext,newtxmath}

\usepackage[T1]{fontenc}
\usepackage{ae,aecompl}

\usepackage{graphicx}	
\usepackage{amsmath}	
\usepackage{amssymb}	
\usepackage{color}
\usepackage{comment}
\usepackage{multirow}

\title[Detectability with 21cm-LAE cross-correlation]{Detectability of 21cm-signal during the Epoch of Reionization with 21cm-Lyman-$\alpha$ emitter\\ cross-correlation. I.}
\author[K. Kubota, S. Yoshiura, K. Takahashi, K. Hasegawa et al.]{
 Kenji Kubota$^1$\thanks{E-mail:175d9001@st.kumamoto-u.ac.jp},
 Shintaro Yoshiura$^1$,
 Keitaro Takahashi$^1$,
 Kenji Hasegawa$^2$,
  \newauthor
 Hidenobu Yajima$^3$,
 Masami Ouchi$^{4,5}$,
 B. Pindor$^{6,7}$,
 and R. L. Webster$^{6,7}$
\\
$^{1}$Faculty of Science, Kumamoto University, 2-39-1 Kurokami, Kumamoto 860-8555, Japan\\
$^{2}$Department of Physics and Astrophysics, Nagoya University Furo-cho, Chikusa-ku, Nagoya, Aichi 464-8602, Japan\\
$^{3}$Frontier Research Institute for Interdisciplinary Sciences, Tohoku University, Sendai 980-8578, Japan\\
$^{4}$Institute for Cosmic Ray Research, The University of Tokyo, 5-1-5 Kashiwanoha, Kashiwa, Chiba 277-8582, Japan\\
$^{5}$Kavli Institute for the Physics and Mathematics of the Universe (WPI), The University of Tokyo,\\ 5-1-5 Kashiwanoha, Kashiwa, Chiba 277-8583, Japan\\
$^{6}$ARC Centre of Excellence for All-sky Astrophysics (CAASTRO)\\
$^{7}$School of Physics, The University of Melbourne, Parkville, VIC 3010, Australia\\
}

\date{Accepted XXX. Received YYY; in original form ZZZ}

\pubyear{2017}

\begin{document}
\label{firstpage}
\pagerange{\pageref{firstpage}--\pageref{lastpage}}
\maketitle

\begin{abstract}
Detection of the redshifted 21cm-line signal from neutral hydrogen in the intergalactic medium (IGM) during the Epoch of Reionization (EoR) is complicated by intense foregrounds such as galactic synchrotron and extragalactic radio galaxies. The 21cm-Lyman-$\alpha$ emitter(LAE) cross-correlation is one of the tools available to reduce the foreground effects because the foreground emission from such radio sources is statistically independent of LAE distribution. LAE surveys during the EoR at redshifts $z=6.6$ and $7.3$ are ongoing by the Subaru Hyper Suprime-Cam (HSC). Additionally, Prime Focus Spectrograph (PFS) will provide precise redshift information of the LAEs discovered by the HSC survey. In this paper, we investigate the detectability of the 21cm signal with the 21cm-LAE cross-correlation by using our improved reionization simulations. We also focus on the error budget and evaluate it quantitatively in order to consider a strategy to improve the signal-to-noise ratio. In addition, we explore an expansion of the LAE survey to suggest optimal survey parameters and show a potential to measure a characteristic size of ionized bubbles via the turnover scale of the cross-power spectrum. As a result, we find that the Murchison Widefield Array (MWA) has ability to detect the cross-power spectrum signal on large scales by combining LAE Deep field survey of HSC. We also show that the sensitivity is improved dramatically at small scales by adding redshift information from the PFS measurements. The Square Kilometre Array (SKA) has a potential to measure the turnover scale with an accuracy of $6\times10^{-3}~{\rm Mpc^{-1}}$.
\end{abstract}



\section{Introduction}

After the Dark Ages, the neutral hydrogen in the IGM was reionized by massive stars and galaxies which emit UV and X-ray photons. This phase of the universe is called the Epoch of Reionization (EoR) and has attracted much attention in communities of both astrophysics and cosmology \citep{2006PhR...433..181F,2012RPPh...75h6901P}. So far, the EoR has often been studied by the Gunn-Peterson test \citep{1965ApJ...142.1633G} in the spectra of high-z quasars, which indicates that the reionization was completed by $z \approx 6$ \citep{2006AJ....132..117F}. On the other hand, the integrated Thomson scattering optical depth of CMB photons implies a redshift of $z \sim 8.8$ in the case of an instantaneous reionization history \citep{2015arXiv150201589P}. However, we have poor information on the early stage of the reionization and the nature of ionizing sources.

The 21cm-line emission from the intergalactic neutral hydrogen is expected to be an effective method to investigate the details of the EoR. Currently, several telescopes are working for this purpose: the Murchison Widefield Array (MWA) \citep{2009IEEEP..97.1497L,2013PASA...30....7T,2013MNRAS.429L...5B}, the LOw Frequency ARray (LOFAR) \citep{2013A&A...556A...2V,2013MNRAS.435..460J} and the Precision Array for Probing the Epoch of Reionization (PAPER) \citep{Jacobs2015,2015ApJ...809...61A}. Although their sensitivities will not be enough to obtain images of the neutral hydrogen distribution, they are sensitive enough to probe its statistical features if we consider only the thermal noise. For the imaging, much higher sensitivity is required and the SKA can be the ultimate telescope for this purpose \citep{2015aska.confE.171C}.

In order to study the statistical feature of 21cm-line signal, the power spectrum and the variance of the probability distribution function (PDF) have been investigated by (semi-)numerical models of reionization \citep{2006PhR...433..181F,2007MNRAS.376.1680P,2008ApJ...689....1S,2010A&A...523A...4B,2013MNRAS.431..621M,2014ApJ...782...66P,2006ApJ...653..815M,2010MNRAS.405.2492H,2015MNRAS.449.4246G,2014MNRAS.443.1113P,2015PhRvD..91l3011D}. The bispectrum and the skewness of the PDF are also fundamental statistical quantities that characterize the fluctuations \citep{2015arXiv150701335S,2015MNRAS.451..467S,2016PASJ...68...61K}. 

However, the statistical detection of 21cm signal suffers from foreground contamination such as galactic synchrotron and extragalactic synchrotron. The 21cm-line signal is typically $O(1)~{\rm mK}$, while the foreground emission is larger by four or more orders of magnitude. Therefore, it is very challenging to identify the EoR signal in the sea of foregrounds. In order to overcome the foregrounds we focus on the cross-correlation between 21cm-line signal and galaxy distribution. Because the foregrounds are statistically independent from the galaxy distribution, the foregrounds do not contribute to the average value of cross-correlation measurements, while they do contribute to the variance.

If galaxies are main sources of ionizing photons, the regions around galaxies are firstly ionized and the ionized bubbles, which are dark in 21cm-line, are formed around the host galaxies. The outside of the ionized bubbles is still partially neutral. On the other hand, neutral regions, which are far from galaxies, are bright in 21cm-line. Thus, negative correlation is expected between 21cm-line signal and galaxy distribution.

In this paper, we focus on the cross-correlation between the 21cm signal and Lyman-$\alpha$ emitters (LAEs). LAEs are high-$z$ galaxies with a strong emission line at the wavelength of 1216 $\rm \AA$. They have been detected by Subaru, Keck, and HST. More than one thousand LAEs have been discovered to date. The farthest LAEs are located at $z \sim 8.7$ \citep{2015ApJ...810L..12Z} and currently, 207 LAEs have been detected at $z = 6.45-6.65$ \citep{2010ApJ...723..869O} and 7 LAEs at $z=7.3$ \citep{2014ApJ...797...16K}. Further, SILVERRUSH project(\citealt{2017arXiv170407455O}, \citealt{2017arXiv170408140S}, \citealt{2017arXiv170500733S}, \citealt{2017arXiv170501222K}) has recently reported 2354 LAEs at $z=5.7$ and $6.6$ as an initial result of ongoing LAE surveys by Hyper Suprime-Cam (HSC) on Subaru telescope. An LAE survey at $z=7.3$ is also being performed and will allow us to probe the epoch before the completion of reionization. In addition to HSC, Prime Focus Spectrograph (PFS), which is under development, is a spectrograph system which can determine the precise redshifts of LAEs through follow-up observations.

Previous studies \citep{2009ApJ...690..252L,2013MNRAS.432.2615W,2014MNRAS.438.2474P,2016MNRAS.459.2741S,2016MNRAS.457..666V,2016arXiv160501734H,2016arXiv161109682H,2017arXiv170107005F} have investigated the detectability of the cross-correlation signal. For example, \citet{2014MNRAS.438.2474P} estimated observational errors on the cross-correlation coefficient and indicated that the cross-power spectrum could be detected under the specifications of the MWA combined with the galaxy survey with redshift errors $\sigma_z \lesssim 0.1$. Further, \citet{2016MNRAS.459.2741S} showed that 1,000 hours observations with the LOFAR would be able to distinguish a fully ionized state from a half ionized state at scales of 3-10 Mpc by using the cross-correlation function. They further found that the SKA1-LOW array will have the potential to distinguish a fully ionized state from quarter ionized state. However, the previous studies used relatively simple EoR model with simulations parameterized the recombination rate and clumping factor in the calculation of ionization structure. Moreover, they have not studied the power of precise redshift determination possible with PFS.

In this paper, we improve the calculation of the cross-correlation signal with numerical simulations taking account of sub-grid effects such as the dependence on halo mass of the recombination rate and the clumping factor of the IGM. Our simulation is consistent with observations of star formation rate density, neutral fraction at redshift $z \sim 6-7$ and the optical depth of cosmic microwave background. Firstly, we confirm the basic features of the cross-correlation and explore the redshift evolution of the cross-correlation. Secondly, we show the detectability of the cross-power spectrum by combining 21cm observation by MWA or SKA with LAE surveys by HSC with and without follow-up of PFS. Thirdly, we study the error budget quantitatively to understand the behavior of sensitivity curves. Finally, we investigate the dependence on survey area and survey depth in LAE survey to suggest optimal survey strategy.

The paper is organized as follows: In Sec.2 we establish notation of the 21cm-LAE cross-correlation such as cross-power spectrum, cross-correlation function, and cross-correlation coefficient. In Sec.3 we describe our numerical simulation for the reionization structure and LAE distribution. In Sec.4, we describe the formalism to estimate the observation errors and sample variance. The basic features and redshift evolution of the cross-correlation signal and our main results concerning the detectability of cross-correlation signal are presented in Sec.5. Finally, we summarize and discuss our results in Sec.6.

\section{21cm-LAE cross-correlation statistics}

The observable quantity of the redshifted 21cm-line is brightness temperature $\delta T_b$ which is determined by the neutral hydrogen fraction $x_{\rm \ion{H}{i}}$ and the matter density fluctuation $\delta_{\rm m}$ as \citep{2006PhR...433..181F},
\begin{equation}
\delta T_b(z) \approx 27 x_{\rm \ion{H}{i}} (1+\delta_{\rm m}) \left( \frac{1+z}{10} \frac{0.15}{\Omega_{\rm m} h^2} \right)^{\frac{1}{2}} \left( \frac{\Omega_{\rm b} h^2}{0.023} \right)~[\rm mK],
\label{dtb}
\end{equation}
where $\Omega_{\rm m}$ and $\Omega_{\rm b}$ are density parameters of matter and baryon, respectively, and $h$ is the Hubble constant in units of $100~{\rm km\ s^{-1}Mpc^{-1}}$. Here, we consider the late stage of EoR so that we assume the spin temperature is much higher than the CMB temperature.

In order to define the 21cm-LAE cross-power spectrum we define spatial fluctuation of $\delta T_b$ as,
\begin{equation}
\delta_{21}({\bf x},z)\equiv\frac{\delta T_b({\bf x},z)-\overline{\delta T_b}(z)}{\overline{\delta T_b}(z)},
\end{equation}
where $\overline{\delta T_b}(z)$ is the spatial average of $\delta T_b$. Similarly, we define fluctuations in galaxy (LAE) abundance as,
\begin{equation}
\delta_{\rm gal}({\bf x},z) \equiv \frac{n_{\rm gal}({\bf x},z)-\bar n_{\rm gal}(z)}{\bar n_{\rm gal}(z)},
\end{equation}
where $n_{\rm gal}({\bf x},z)$ is the number density of galaxies (LAEs) and $\bar n_{\rm gal}(z)$ is the spatial average of $n_{\rm gal}$. Note that both $\delta_{21}({\bf x})$ and $\delta_{\rm gal}({\bf x})$ are dimensionless quantities. Defining $\tilde \delta_{21}({\bf k})$ and $\tilde \delta_{\rm gal}({\bf k})$ to be Fourier transform of $\delta_{21}({\bf x})$ and $\delta_{\rm gal}({\bf x})$, respectively, the cross-power spectrum $P_{\rm 21,gal}({\bf k})$ is given by
\begin{equation}
\langle \tilde\delta_{21}({\bf k_1}) \tilde\delta_{\rm gal}({\bf k_2}) \rangle \equiv (2\pi)^3 \delta_D({\bf k_1+k_2}) P_{\rm 21,gal}({\bf k_1}),
\end{equation}
where $\delta_D({\bf k})$ is the Dirac delta function. The dimensionless cross-power spectrum is given by
\begin{equation}
\Delta_{\rm 21,gal}^2(k) = \frac{k^3}{2\pi^2} P_{\rm 21,gal}(k).
\end{equation}

The cross-correlation function $\xi_{\rm 21,gal}({\bf r})$ is defined as,
\begin{equation}
\xi_{\rm 21,gal}({\bf r}) \equiv \langle \delta_{21}({\bf x}) \delta_{\rm gal}({\bf x+r})\rangle,
\end{equation}
which is related to the cross-power spectrum by Fourier transform:
\begin{equation}
\xi_{\rm 21,gal}(r) = \frac{1}{(2\pi)^3} \int P_{\rm 21,gal}(k) \frac{\sin(kr)}{kr} 4\pi k^2 dk.
\end{equation}

Finally, the cross-correlation coefficient is defined as,
\begin{equation}
r_{\rm 21,gal}(k) = \frac{P_{\rm 21,gal}(k)}{\sqrt{P_{21}(k)P_{\rm gal}(k)}}.
\end{equation}
where $P_{21}(k)$ and $P_{\rm gal}(k)$ are auto-power spectra of 21cm-line brightness temperature and galaxies, respectively, given by,
\begin{eqnarray}
\langle \tilde\delta_{21}({\bf k_1})\tilde\delta_{21}({\bf k_2})\rangle\equiv(2\pi)^3\delta_D({\bf k_1+k_2})P_{21}({\bf k_1}),\\
\langle \tilde\delta_{\rm gal}({\bf k_1})\tilde\delta_{\rm gal}({\bf k_2})\rangle\equiv(2\pi)^3\delta_D({\bf k_1+k_2})P_{\rm gal}({\bf k_1}).
\end{eqnarray}

\section{Simulation data}

We compute the cross-correlation signal using our numerical simulations. In this section, we describe how the simulate the reionization process and obtain mock LAE samples. More details will be presented elsewhere (Hasegawa et al. in preparation).  

\subsection{Reionization model}

Previous radiation hydrodynamics (RHD) simulations have shown that radiative feedback regulates star formation rates in galaxies and the IGM clumping factor during the EoR \citep{Pawlik09, Finlator12, Wise12, 2013MNRAS.428..154H}. 
However, due to expensive computational costs, it is very difficult to conduct cosmological RHD simulations with a large enough volume to sample the large-scale ionization structure of the IGM and high enough spatial resolution to resolve radiative feedback on galaxies. 
Hence in our reionization simulations, we first constructed sub-grid models of ionizing sources and IGM clumping factor from a cosmological RHD simulation with high resolution, and then use the models for post-processing radiative transfer calculation (\cite{2016arXiv160301961H}, Hasegawa et al. in preparation). 

The RHD simulation used for deriving the sub-grid models was performed with $2\times512^3$ particles in a simulation volume of $(20~\rm Mpc)^3$. 
We adopted an RHD method similar to that in \citet{2013MNRAS.428..154H}. 
Since the escape fraction is sensitive to the amount and distribution of gas in galaxies, the escape fraction in the RHD simulation is regulated by UV and supernovae feedback effects and turns out to be high for less massive galaxies.
Additionally, the RHD simulation showed that the clumping factor varies not only with the local ionization degree but also with the local density \citep{2016arXiv160301961H}. 
To appropriately consider these remarkable features found in the RHD simulation, we made look-up tables for the spectral energy distribution (SED) of galaxies (as a two-dimensional function of the halo mass and the local ionization degree) and the IGM clumping factor (as a two-dimensional function of the local IGM density and the local ionization degree) from the RHD simulation results. 
The stellar age dependent SED was computed with $\rm P\acute{E}GASE2$\footnote{http://www2.iap.fr/users/fioc/PEGASE.html} \citep{PEGASE}, assuming the Salpeter mass function ranging from $0.1~M_\odot$-$120~M_\odot$.
With this environment-dependent clumping factor model, the clumping factor tends to be higher as the local density increases and as the local ionization degree decreases. The clumping factor ranges from $\sim 1$ to $\sim 100$. Similarly to previous studies \citep{Pawlik09,Finlator12}, the typical value corresponds $\sim 3$ in highly ionized regions (see Figure 2 of \citealt{2016arXiv160301961H}).

The matter distribution at each redshift is obtained from a large-scale $N$-body simulation performed with a massive parallel TreePM code GreeM\footnote{http://hpc.imit.chiba-u.jp/\~ishiymtm/greem/} \citep{Ishiyama2009b, Ishiyama2012}, for which $4096^3$ particles in a $(160~\rm Mpc)^3$ box are utilized. 
We divide the whole volume into $256^3$ cells for the post-processing radiative transfer calculation and thus each grid size corresponds to 0.625 Mpc on a side. 
The time evolution of the \ion{H}{i}, \ion{He}{i}, and \ion{He}{ii} fractions ($x_{\rm \ion{H}{i}}$, $x_{\rm \ion{He}{i}}$ and $x_{\rm \ion{He}{ii}}$) at each position is given by 
\begin{eqnarray}
		\frac{dx_{\rm \ion{H}{i}}}{dt} &=& -k^{\rm \ion{H}{i}}_{\gamma} - k_{\rm c}^{\rm \ion{H}{i}} x_{\rm \ion{H}{i}}n_{\rm e} + 
		C\alpha_{\rm B}^{\rm \ion{H}{ii}} x_{\rm \ion{H}{ii}} n_{\rm e}, \\
		\frac{dx_{\rm \ion{He}{i}}}{dt} &= &-k^{\rm \ion{He}{i}}_{\gamma} - k^{\rm \ion{He}{i}}_{\rm c} x_{\rm \ion{He}{i}}n_{\rm e}+
		C\alpha_{\rm B}^{\rm \ion{He}{ii}}  x_{\rm \ion{He}{ii}} n_{\rm e}, \\
		\frac{dx_{\rm \ion{He}{ii}}}{dt} &=& k^{\rm \ion{He}{i}}_{\gamma} -k^{\rm \ion{He}{ii}}_{\gamma}  
		+ k^{\rm \ion{He}{i}}_{\rm c} x_{\rm \ion{He}{i}} n_{\rm e} 
		- k^{\rm \ion{He}{ii}}_{\rm c} x_{\rm \ion{He}{ii}} n_{\rm e} \nonumber \\
		&&-C\alpha_{\rm B}^{\rm \ion{He}{ii}} x_{\rm \ion{He}{ii}}n_{\rm e} + C\alpha_{\rm B}^{\rm \ion{He}{iii}} x_{\rm \ion{He}{iii}}n_{\rm e}, 
\end{eqnarray}
where $k^i_{\gamma}$, $k^i_{\rm c}$, and $\alpha_ {\rm B}^i$ are the photo-ionization, collisional ionization and case-B recombination rates for $i$-th species, respectively.  
Here, $C$ is the environment-dependent clumping factor mentioned above. 
The photo-ionization rates at a given position are obtained by solving radiative transfer, and described as 
\begin{equation}
		k^i_{\gamma} = \sum_j \frac{x_{i}}{4\pi R_j^2} \int^\infty_{\nu_i} 
		\frac{L_{\nu,j}}{h\nu} \sigma_{i}(\nu) {\rm e}^{-\tau_{\nu,j}} d\nu, 
\end{equation}
where $\sigma_i(\nu)$ is the cross section for $i$-th species, $\nu_i$ is the Lyman limit frequency of $i$-th species. 
The subscript $j$ indicates the index of an ionizing source, $R_j$ and $\tau_{\nu,j}$ are respectively the distance and the optical depth from $j$-th ionizing source.  
The SED of $j$-th ionizing source is determined by referring to the look-up table of SED. 
Thermal evolution at each position obeys the following equation;  
\begin{equation}
	\frac{dT_{\rm g}}{dt} = (\gamma-1)\frac{\mu m_{\rm p}}{k_{\rm B}\rho}
	\left(\frac{k_{\rm B}T_{\rm g}}{\mu m_{\rm p}}\frac{d\rho}{dt} + \Gamma - \Lambda \right) 
	- \mu T_{\rm g}\frac{d}{dt}\left(\frac{1}{\mu}\right),
\end{equation}
where $\gamma$, $m_{\rm p}$, $\mu$, $\rho$, and $k_{\rm B}$ are the adiabatic index, the proton mass, mean molecular weight, gas mass density, and the Boltzmann constant, respectively. 
The \ion{H}{i}, \ion{He}{i}, and \ion{He}{ii} photo-ionization processes contribute to the heating rate $\Gamma$. 
Each contribution is written as 
\begin{equation}
	\Gamma_{i, \gamma} = \sum_j \frac{n_{i}}{4\pi R_j^2} \int^\infty_{\nu_i} 
	\frac{L_{\nu,j}}{h\nu} (h\nu-h\nu_i)\sigma_{i}(\nu) {\rm e}^{-\tau_{\nu,j}} d\nu.  
\end{equation}
During the post-processing radiative transfer calculation, $L_{\nu,j}$ and $C(\bf x)$ are estimated from the look-up tables, referring to the halo mass, the local IGM density, and the local ionization degree. 

\begin{figure}
\begin{center}
\includegraphics[width=9cm,trim=1.5cm 0cm 1cm 0cm]{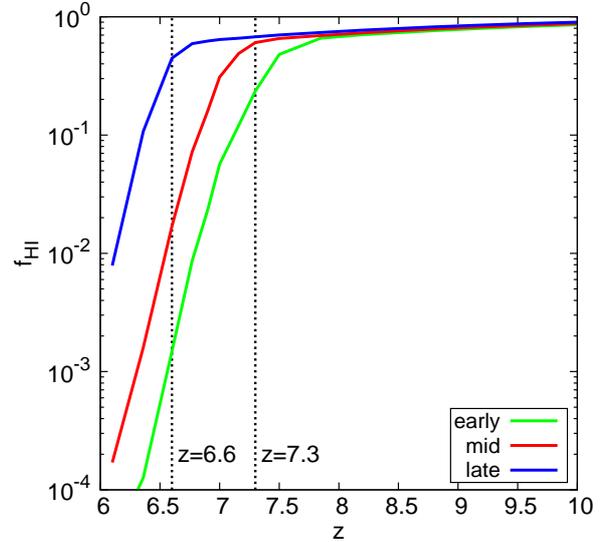}
\end{center}
\caption{Evolution of the mean neutral hydrogen fraction $f_{\rm HI}$ in our reionization simulation box as a function of redshift. The green, red, blue lines show the evolution in the early, mid, and late model, respectively.}
\label{fig:f_HI}
\end{figure}

Other than the fiducial model, we perform two additional reionization simulations with different ionizing photon production rate models. 
The ionizing photon production rates in the additional two runs are set to be 1.5 times higher or lower than that in the fiducial model.
We refer to these three models as the late, mid, and early reionization models, respectively. 
These ionizing-source models accurately reproduce the neutral hydrogen fraction at $z\sim6$ indicated by QSO spectra and the Thomson scattering optical depth for the CMB photons, simultaneously.
Fig.~\ref{fig:f_HI} shows the evolution of the mean neutral hydrogen fraction of the three simulations.
The optical depths are 0.0552, 0.0591, 0.0648 for the late, mid, and early models, respectively, while the Planck observation gives $0.066 \pm 0.016$\citep{2015arXiv150201589P}.

\begin{figure}
\begin{center}
\includegraphics[width=5cm,trim=4cm 0cm 4cm 0cm]{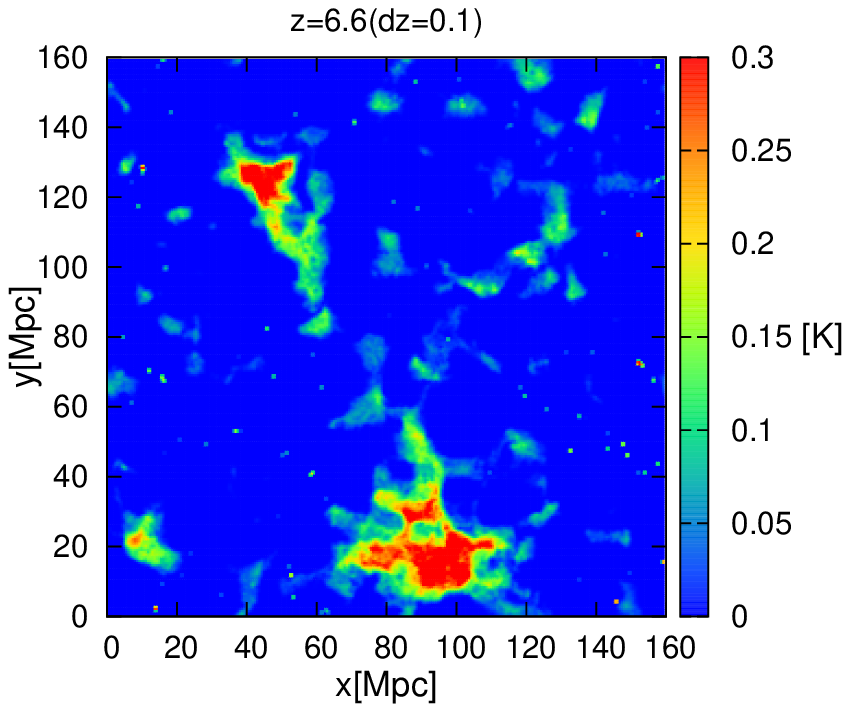}
\includegraphics[width=8.5cm,trim=1.3cm 0cm 0.5cm 0cm]{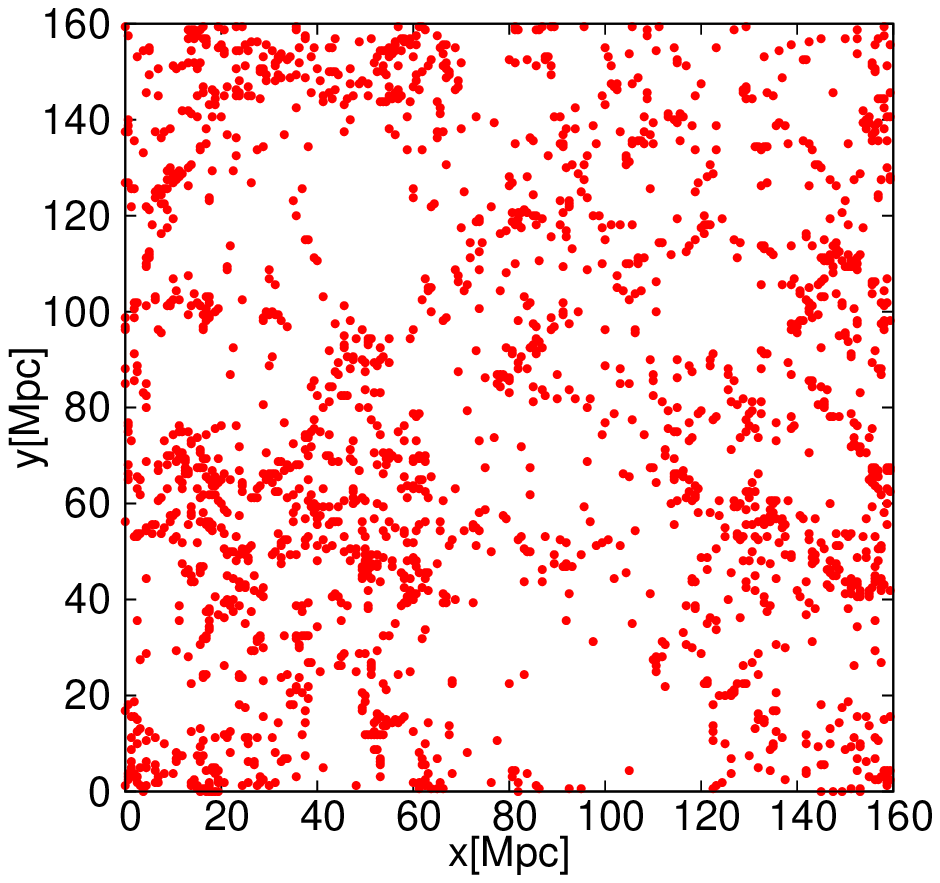}
\end{center}
\caption{{\bf Top}: the 21cm brightness temperature in mid model at redshift $z=6.6$. In fully ionized region $\delta T_b\sim 0\rm mK$. {\bf Bottom}: the associated LAE distribution. The panels are maps integrated within $\Delta z=0.1 \sim 40 \rm Mpc$.}
\label{fig:dTb_LAE_mid}
\end{figure}

We finally evaluate the differential brightness temperature $\delta T_b$ from Eq.~(\ref{dtb}), assuming that the spin temperature $T_S$ is fully coupled with the gas temperature $T_{\rm g}$. 
We note that this assumption is valid as far as we focus on the later stage of the EoR \citep{Baek09}. 
The map of $\delta T_b$ at $z=6.6$ in the mid model is shown in the top panel of Fig.~\ref{fig:dTb_LAE_mid}. 

\subsection{Galaxy (LAE) model}

The mock LAE samples are obtained via two steps. 
Firstly, we determine the Ly$\alpha$ luminosity of each galaxy. 
Next, we evaluate the Ly$\alpha$ transmission rate through the simulated IGM for each galaxy by integrating the Ly$\alpha$ optical depth along a given direction. 
Since the ionization structure in each galaxy is calculated in the RHD simulation described in the previous subsection, we can estimate the intrinsic Ly$\alpha$ luminosity of each galaxy from the RHD simulation results. 
In galaxies, Ly$\alpha$ photons are mainly produced via the recombination process and the collisional excitation process \citep{Yajima12}. 
By counting the number of Ly$\alpha$ photons produced by these two processes, we found that the intrinsic Ly$\alpha$ luminosity $L_{\alpha, \rm int}$ of each galaxy with halo mass being greater than $10^{10}M_{\odot}$ is roughly expressed as 
\begin{equation}
	L_{\alpha,\rm int}\approx10^{42}\Bigl(\frac{M_{\rm h}}{10^{10}{\rm M_{\odot}}}\Bigr)^{1.1}[\rm erg/s], 
\end{equation}
where $M_{\rm h}$ is the halo mass. 
We note that the dependence on the halo mass is almost identical to that for the star formation rate in the RHD simulation. 
It is usually expected that the intrinsic Ly$\alpha$ photons are absorbed by interstellar dust during the numerous scattering events. 
In this paper, we treat the fraction of Ly$\alpha$ photons escaping from a galaxy, $f_{\rm esc, \alpha}$, as a free parameter, because the absorption of Ly$\alpha$ photons by dust grains is not taken into account in the RHD simulation. It should be noted that we ignore the dispersion of Ly$\alpha$ luminosity for a given halo mass. \citet{2018arXiv180100067I} have shown that the absence of the dispersion in $L_{\alpha,\rm int}$ leads to a stronger clustering of LAEs than observed one at small scales.

The Ly$\alpha$ flux is further attenuated by neutral hydrogen in the IGM before it can be observed. 
It is essential to determine the Ly$\alpha$ line profile emerging from the surface of a galaxy for evaluating the fraction of the Ly$\alpha$ flux transmitted through the IGM, because the Ly$\alpha$ transmission rate is sensitive to the line profile. 
In this work, we use the line profiles obtained by solving Ly$\alpha$ radiative transfer with an expanding spherical cloud model in which the radial velocity is assumed to obey $v(r) = V_{\rm out}\left(\frac{r}{r_{\rm vir}}\right)$, where $r_{\rm vir}$ and $V_{\rm out}$ are the virial radius of a halo and the galactic wind velocity \citep{Yajima17}. 
The line profile is controlled by two parameters; the galactic wind velocity $V_{\rm out}$ and the \ion{H}{i} column density in a galaxy $N_{\rm \ion{H}{i}}$. 
In the expanding cloud model, photons with short wavelengths are selectively scattered by outflowing gas. 
As a result, an asymmetric profile with a characteristic peak at a wavelength longer than 1216~\AA~emerges from the surface of a galaxy. 

Using the obtained line profile $\phi_{\alpha}(\nu)$, the Ly$\alpha$ transmission rate $T_{\alpha, \rm IGM}$ is calculated as 
\begin{equation}
	T_{\alpha, \rm IGM} = 
	\frac{\int \phi_{\alpha}(\nu _0)~e^{-\tau_{{\nu_0},\rm IGM}} d\nu_0}
	{\int \phi_{\alpha}(\nu_0) d\nu_0}\,,
\end{equation}
where $\nu_0$ is the frequency in the rest-frame of a galaxy, $\tau_{\nu,\rm IGM}$ is the optical depth through the IGM described as 
\begin{equation}
	\tau_{\nu_0, \rm IGM} = \int_{r_{\rm vir}}^{l_{\rm p,max}} 
	s_\alpha(\nu,T_{\rm g}) n_{\rm \ion{H}{i}} dl_{\rm p}, 
\end{equation}
where $s_{\alpha}$ is the Ly$\alpha$ cross section of neutral hydrogen. Note that the frequency in the rest frame of the expanding gas, $\nu$, is given by 
\begin{equation}
	\nu = \nu_0\left(1-\frac{H(z)l_{\rm p}}{c} \right), 
\end{equation}
where $l_{\rm p}$ is the distance from an LAE candidate in the physical coordinate. 
The upper bound of the integration, $l_{\rm p, \rm max}$, is set to be 80 comoving Mpc. 
The Ly$\alpha$ transmission rate $T_{\alpha, \rm IGM}$ tends to be higher as the outflow velocity $V_{\rm out}$ or the \ion{H}{i} column density $N_{\rm \ion{H}{i}}$ increases, because the remarkable peak shifts towards redder wavelengths \citep{Yajima17}. 

\begin{figure}
\begin{center}
\includegraphics[width=9cm,trim=1.5cm 0cm 1cm 0cm]{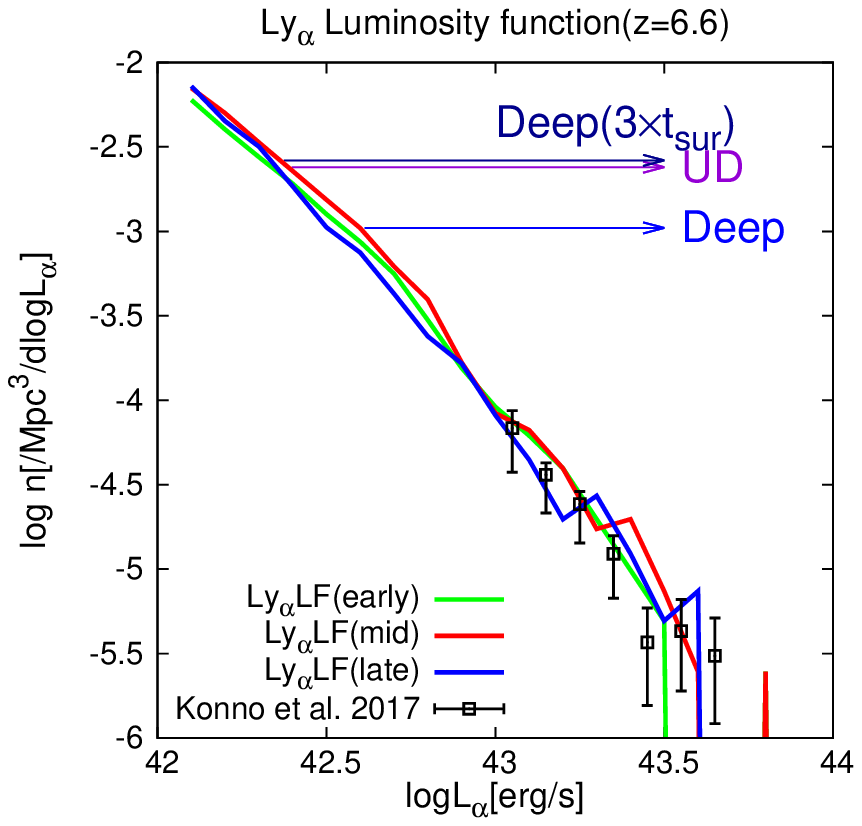}
\includegraphics[width=9cm,trim=1.5cm 0cm 1cm 0cm]{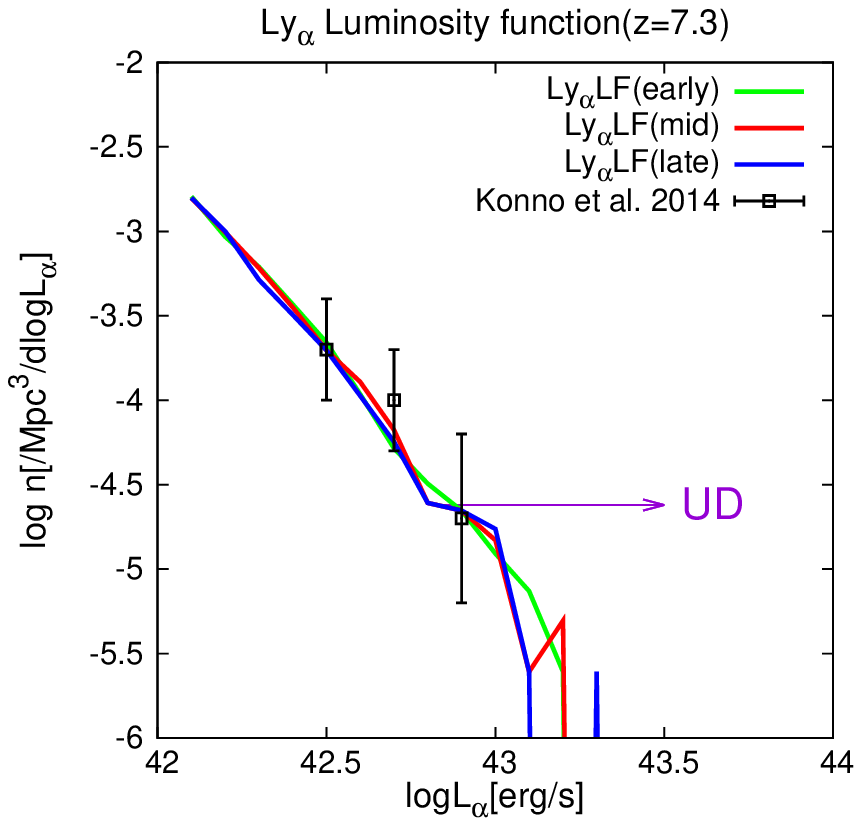}
\end{center}
\caption{Simulated Ly$\alpha$ luminosity function and observed LF at redshift $z=6.6$ (top) and $z=7.3$ (bottom). The green, red, and blue solid lines show the simulated LFs in the early, mid, late model, respectively. In the top panel, the arrows represent the detectable luminosity range in Ultra-deep, Deep field, and the case of $3 \times t_{\rm sur}$ in Deep field of HSC LAE surveys.}
\label{fig:LF}
\end{figure}

In summary, observable Ly$\alpha$ luminosity is given by 
\begin{equation}
	L_{\alpha, \rm obs} = f_{\rm esc, \alpha}T_{\alpha, \rm IGM}L_{\alpha, \rm int}. 
\end{equation}
As described above, the transmission rate $T_{\alpha, \rm IGM}$ implicitly depends on $V_{\rm out}$ and $N_{\rm \ion{H}{i}}$. 
Thus, the observable Ly$\alpha$ luminosity is determined not only by the neutral hydrogen distribution in the IGM, but also three parameters, i.e., $f_{\rm esc, \alpha}$, $V_{\rm out}$ and $N_{\rm \ion{H}{i}}$. 
In this work, we set the parameters to be $0.16\leq f_{\rm esc,\alpha}\leq0.45$, $V_{\rm out}=150{\rm km/s}$, $N_{\rm \ion{H}{i}}=10^{19}$ or $10^{20}\rm cm^{-2}$ so that simulated Ly$\alpha$ luminosity functions match the observed LFs.
The parameters we set are summarized in Table\ref{tb:parameter}.
Fig.\ref{fig:LF} shows the comparison between the simulated Ly$\alpha$ luminosity functions with the chosen parameters and observed LFs at redshifts $z=6.6$\citep{2017arXiv170501222K} and $z=7.3$\citep{2014ApJ...797...16K}.
Although the simulated LFs are well consistent with observations, as mentioned above, our simple LAE model cannot reproduce clustering properties of LAEs provided by recent observation with HSC \citep{2017arXiv170407455O,2018arXiv180100067I}. 
We note that LAE bias in our LAE model is larger than the results in \citet{2017arXiv170407455O,2018arXiv180100067I} by one order of magnitude at $k\sim1.0{\rm Mpc}^{-1}$. This inconsistency will diminish the power of the cross-spectrum signal by the magnitude and possibly affects the detectability of the cross-power signals on small scales. We will discuss this point in the future work.

The bottom panel of Fig. \ref{fig:dTb_LAE_mid} shows the distribution of observable LAEs ($L_{\alpha, \rm obs} >10^{42}\rm erg/s$) in the mid model at $z=6.6$. 
The comparison between the 21cm and LAE maps indicates that LAEs clearly reside in the ionized region ($\delta T_b\sim0$\rm mK) and the 21cm brightness temperature is high in the no LAEs region. 
This anti-correlation was seen in the previous works.

\begin{table}
\begin{center}
\caption{Parameter sets we chose in our LAE model at redshift $z=6.6$ and $7.3$. We choose $N_{\rm \ion{H}{i}}=10^{19}~\rm{cm}^{-2}$ at redshift $z=6.6$ and $10^{20}~\rm{cm}^{-2}$ at redshift $z=7.3$. The LAE models in the early, mid, late model are set by adjusting $f_{\rm esc,\alpha}$.}
\renewcommand{\arraystretch}{1.3}
\begin{tabular}{|c|c|c|c|c|} \hline
$z$ & model & $f_{\rm esc,\alpha}$ &  $V_{\rm out}$[km/s] & $N_{\rm \ion{H}{i}}$[$\rm{cm}^{-2}$] \\ \hline \hline
           &early  & 0.22       & 150     &$10^{19}$  \\ \cline{2-5}
  6.6    &mid    & 0.25       & 150     &$10^{19}$  \\ \cline{2-5}
           &late    & 0.45       & 150     &$10^{19}$  \\ \cline{1-5}
           &early  & 0.16       & 150     &$10^{20}$   \\ \cline{2-5} 
  7.3    &mid    & 0.30       & 150     &$10^{20}$   \\ \cline{2-5}
           &late    & 0.37       & 150     &$10^{20}$   \\[-3pt] \hline

\end{tabular}
\label{tb:parameter}
\end{center}
\end{table}


\section{Detectability}

In this section, we describe how to estimate the error on the cross-power spectrum.
We calculate the error according to \citet{2009ApJ...690..252L,2007ApJ...660.1030F}. As to observation facilities, we consider combining the 21cm-line observation by the MWA and SKA with the LAE survey by Subaru HSC and follow-up observations by PFS.

\subsection{Statistical error}

First of all, we account for enhancement of the power spectrum by redshift space distortion as $P(k,\mu) = (1 + \beta \mu^2)^2 P(k)$, where $\mu$ is the cosine of the angle between $\bf k$ and the line-of-sight. $\beta = \Omega_m^{0.6}(z)/b$ and $b$ is a bias factor\citep{1987MNRAS.227....1K}. The bias factor is given by $b_{\rm gal}^2(k) = P_{\rm gal}(k)/P_{\rm DM}(k)$ and here we compute this as $b_{\rm gal}^2(k) = P_{\rm gal}(k)/P_{\rm density}(k)$ assuming $P_{\rm density}(k) \approx P_{\rm DM}(k)$, where $P_{\rm DM}(k)$ and $P_{\rm density}(k)$ are dark matter and gas density power spectra, respectively. We also set $b_{21} = 1$ for 21cm-line power spectrum in the error estimation since we do not take into account peculiar velocity in our 21cm-line simulation. However, for simplicity, we neglect the effect for the cross-correlation signal in this paper since the effect is actually negligible.

Without systematic errors, the error on a measurement of the 21cm power spectrum for a particular mode ($k,\mu$) is given by \citep{2006ApJ...653..815M}
\begin{equation}
\delta P_{21}(k,\mu) = P_{21}(k,\mu) + \frac{T_{\rm sys}^2}{Bt_{\rm int}} \frac{D^2\Delta D}{n(k_\perp)} \Bigl( \frac{\lambda^2}{A_e} \Bigr)^2,
\label{eq:thermal}
\end{equation}
where $T_{\rm sys}$ is the system temperature which is estimated as $\sim 280 [(1+z)/7.5]^{2.3}~{\rm K}$. $B$ and $t_{\rm int}$ are the survey bandpass and the integration time for 21cm observation, respectively. $D$ is the comoving distance to the 21cm survey volume and the comoving survey width $\Delta D$ is given by $\Delta D = 1.7 (\frac{B}{0.1\rm MHz}) (\frac{1+z}{10})^{1/2} (\frac{\Omega_m h^2}{0.15})^{-1/2}$. $n(k_{\perp})$ is the number density of baselines in observing the perpendicular component of the wave vector, $k_{\perp} = (1-\mu^2)^{1/2}k$. We assume that it is decreased continuously as $r^{-2}$. $A_e$ is the effective area of each antenna tile and $\lambda$ is the observed 21cm wavelength. The first and second terms represent sample variance and thermal noise, respectively.

Similarly, the error on the galaxy survey for a particular mode is given by \citep{1994ApJ...426...23F,1997PhRvL..79.3806T}
\begin{equation}
\delta P_{\rm gal}(k,\mu) = P_{\rm gal}(k,\mu) + n_{\rm gal}^{-1} \exp({k_{\parallel}^2}\sigma_r^2),
\label{eq:shot}
\end{equation}
where $n_{\rm gal}$ is the mean number density in the galaxy survey. Its inverse approximately is regarded as shot noise; $k_{\parallel}$ is the parallel component of wave number, $k_{\parallel}=\mu k$. $\sigma_r = c \sigma_z/H(z)$ where $\sigma_z$ is the redshift error in the galaxy survey. Here the first term is sample variance and the second term is a product of shot noise and redshift errors.

With the errors on the 21cm observation and the galaxy survey, the error on the cross-power spectrum for a particular mode is give by
\begin{equation}
2 [\delta P^2_{\rm 21,gal}(k,\mu)] = P^2_{\rm 21,gal}(k,\mu) + \delta P_{21}(k,\mu) \delta P_{\rm gal}(k,\mu).
\label{eq:cross-error}
\end{equation}
The first term represents sample variance on the cross-power spectrum and the second term is a product of Eqs.~(\ref{eq:thermal}) and (\ref{eq:shot}). We then compute the error on the cross-power spectrum by summing the errors for each $\bf k$-modes in inverse form. The errors on the spherically averaged cross-power spectrum are,
\begin{equation}
\frac{1}{\delta P^2_{\rm 21,gal}(k)} = \sum_\mu \Delta \mu \frac{\epsilon k^3 V_{\rm sur}}{4\pi^2} \frac{1}{\delta P^2_{\rm 21,gal}(k,\mu)},
\label{eq:error-volume}
\end{equation}
where $\epsilon=\Delta k/k$ is the logarithmic width of the spherical shell, and $V_{\rm sur}$ is the effective survey volume for 21cm radio telescope which is given by $V_{\rm sur} = D^2 \Delta D (\lambda^2/A_e)$. If the galaxy survey has a smaller volume than 21cm-line survey, we set $V_{\rm sur} = V_{\rm gal}$. We note the typical survey volume for the 21cm observation and the galaxy survey are of order $V_{\rm sur}\sim 10^9\rm Mpc^{3}$ and $V_{\rm gal}\sim 10^6\rm Mpc^{3}$, respectively. In our calculation, the survey volume of the 21cm observation is much larger than that of the galaxy survey so that the sensitivity on the cross-power spectrum is limited by $V_{\rm gal}$.

We then calculate the total signal-to-noise (S/N) ratio which is summation of the S/N in each $k$ bin,
\begin{equation}
(S/N)^2_{\rm total} = \sum_{i}^{N_{\rm bin}} \Bigl(\frac{\Delta k}{\epsilon k_i}\Bigr) (S/N)^2_i,
\end{equation}
where $N_{\rm bin}$ and $\Delta k$ are the number of bins and the bin size, respectively.

Next, we consider a case where PFS is not available and precise redshift information of LAEs cannot be obtained. In this case, a 2D projection of 3D cross-correlation signal can be obtained. We can derive observational errors in this case easily from the above 3D case. Since $\mu$ is always zero ($k_{\parallel}=0$) in 2D space, the error on a measurement of the 21cm power spectrum is reduced to:
\begin{equation}
\delta P_{21,{\rm 2D}}(k) = P_{21,{\rm 2D}}(k) + \frac{T_{\rm sys}^2}{Bt_{\rm int}} \frac{D^2}{n(k)} \Bigl( \frac{\lambda^2}{A_e} \Bigr)^2,
\label{eq:thermal_2D}
\end{equation}
where $P_{21,{\rm 2D}}(k)$ is 2D 21cm power spectrum. The error on the galaxy survey is also reduced to:
\begin{equation}
\delta P_{\rm gal,2D}(k) = P_{\rm gal,2D}(k) + n_{\rm gal}^{-1},
\label{eq:shot_2D}
\end{equation}
where $P_{\rm gal,2D}(k)$ is 2D galaxy power spectrum. Moreover, one can reduce the error by integrating the signal within an annulus in Fourier space. The number of samples in annulus is given by:
\begin{equation}
N_{\rm a} = 2\pi k\Delta k \frac{S_{\rm sur}}{(2\pi)^2},
\label{eq:shot_2D}
\end{equation}
where $k=|\bf k|$, and $S_{\rm sur}$ is field of view. Finally, the error for the averaged 2D 21cm-LAE cross-power spectrum is given by:
\begin{equation}
\delta P^2_{\rm 21,gal,2D}(k) = \frac{1}{N_{\rm a}}[P^2_{\rm 21,gal,2D}(k) + \delta P_{21,{\rm 2D}}(k) \delta P_{\rm gal,2D}(k)].
\label{eq:cross-error_2D}
\end{equation}

Later, we will investigate the error budget of cross-correlation measurements, so let us represent Eq.~(\ref{eq:cross-error}) more simply. We denote the thermal noise in Eq.~(\ref{eq:thermal}) as $\sigma_{\rm N}$, the shot noise in Eq.~(\ref{eq:shot}) as $\sigma_{\rm g}$ and the error on the cross-power spectrum as $\sigma_{\rm A}$. Then, Eq.~(\ref{eq:cross-error}) can be rewritten as
\begin{equation}
\sigma_{\rm A}(k)\propto \sqrt{P^2_{21,\rm gal}+P_{21}P_{\rm gal}+P_{21}\sigma_{\rm g}+\sigma_{\rm N}P_{\rm gal}+\sigma_{\rm N}\sigma_{\rm g}}.
\label{eq:budget}
\end{equation}
Each term in Eq.~(\ref{eq:budget}) represents a component of the error on the cross-power spectrum. The error is determined by the 5 terms. We will compare these terms later.

\subsection{MWA and SKA1-low}

With these expressions we describe the specifications for the 21cm observation. The MWA has a large field of view ($\sim 800~{\rm deg}^2$) on the sky and effective area $A_e = 14~{\rm m}^2$ at $z=8$ \citep{Bowman2006}. Each antenna tile is $4~{\rm m}$ wide and the antennas are packed as closely as possible within a compact core out to a maximum baseline of $1.5~{\rm km}$. We assume 256 antenna tiles within $750~{\rm m}$, a survey bandpass of $B = 8~{\rm MHz}$, and 1,000 hrs observing time.

The SKA is a next-generation low-frequency radio telescope that will be operated from 2020. The SKA1-low, the low-frequency component of the SKA, will consist of 670 antenna tiles within $1000~{\rm m}$ with effective area $A_e = 462~{\rm m}^2$ at $z=8$ \citep{2016SPIE.9906E..28W}. The SKA1-low also has a wide field-of-view of $\sim 25~{\rm deg}^2$. As well as the MWA, we assume the packed configuration, a survey bandpass of $B = 8~{\rm MHz}$, and observing time of 1,000 hrs.

\subsection{HSC and PFS}

Hyper Sprime-Cam (HSC) is a huge camera with a wide field-of-view of $1.5~{\rm deg}^2$ for Subaru telescope. Narrow-band LAE surveys with HSC are currently ongoing and have two layers; Ultra-deep field and the Deep field survey. The Ultra-deep field survey has $3.5~{\rm deg}^2$ survey area at redshift $z=6.6$ and $7.3$. It will discover $\sim1700$ and $\sim39$ LAEs with the detection limit of the observed luminosity $L_{\alpha} = 2.5 \times 10^{42}~{\rm erg/s}$ and $6.8 \times 10^{42}~{\rm erg/s}$ at redshift $z=6.6$ and $7.3$, respectively. On the other hand, The Deep field survey has a wider survey area of $\sim 27~{\rm deg}^2$ and a larger detection limit of the observed luminosity $L_{\alpha} = 4.1 \times 10^{42}~{\rm erg/s}$. It will discover $\sim5500$ LAEs at redshift $z=6.6$. Because of systemic redshift uncertainties of narrow-band surveys, the redshift has an uncertainty of order $\Delta z=0.1$, which corresponds to a radial distance of $\sim~{\rm 40Mpc}$. Thus, they provide LAE maps which are integrated within $\Delta z$, where the ionization structure and the associated LAE clustering signature are smeared out and information on $k_{\parallel}$ modes is lost.

Prime Focus Spectrograph (PFS) is a spectrograph system on Subaru telescope and is currently under development. It has a large spectral resolving power of $R\sim 3000$ as well as a wide field-of-view of $\sim 1.3~{\rm deg}^2$. Thus, follow-up observations of HSC fields allow us to determine the precise redshifts of the LAEs discovered by HSC. We calculate the error on the galaxy survey by assuming $\sigma_z = 0.0007$.

\section{Results}

\subsection{Cross-correlation signal}
\begin{figure}
\begin{center}
\includegraphics[width=8cm,trim=0cm 0cm 0cm 0cm]{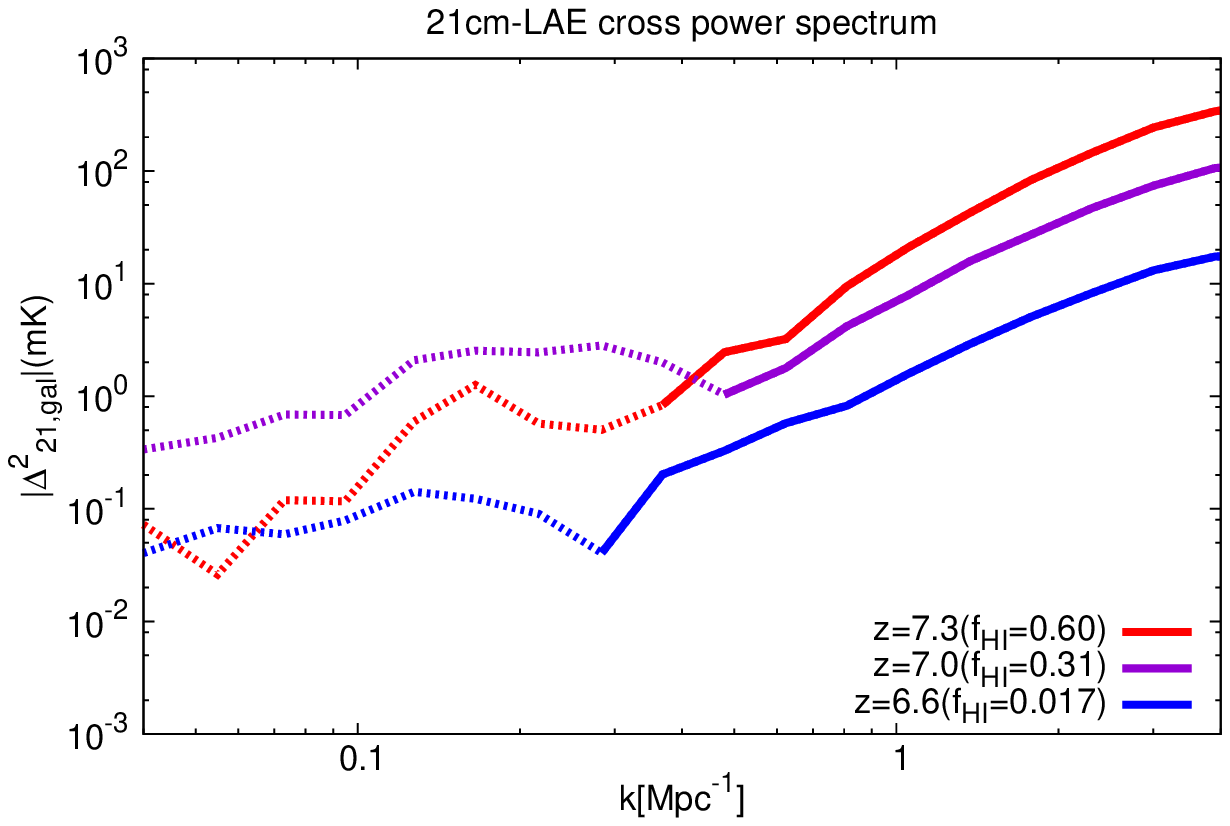}
\includegraphics[width=8cm,trim=0cm 0cm 0cm 0cm]{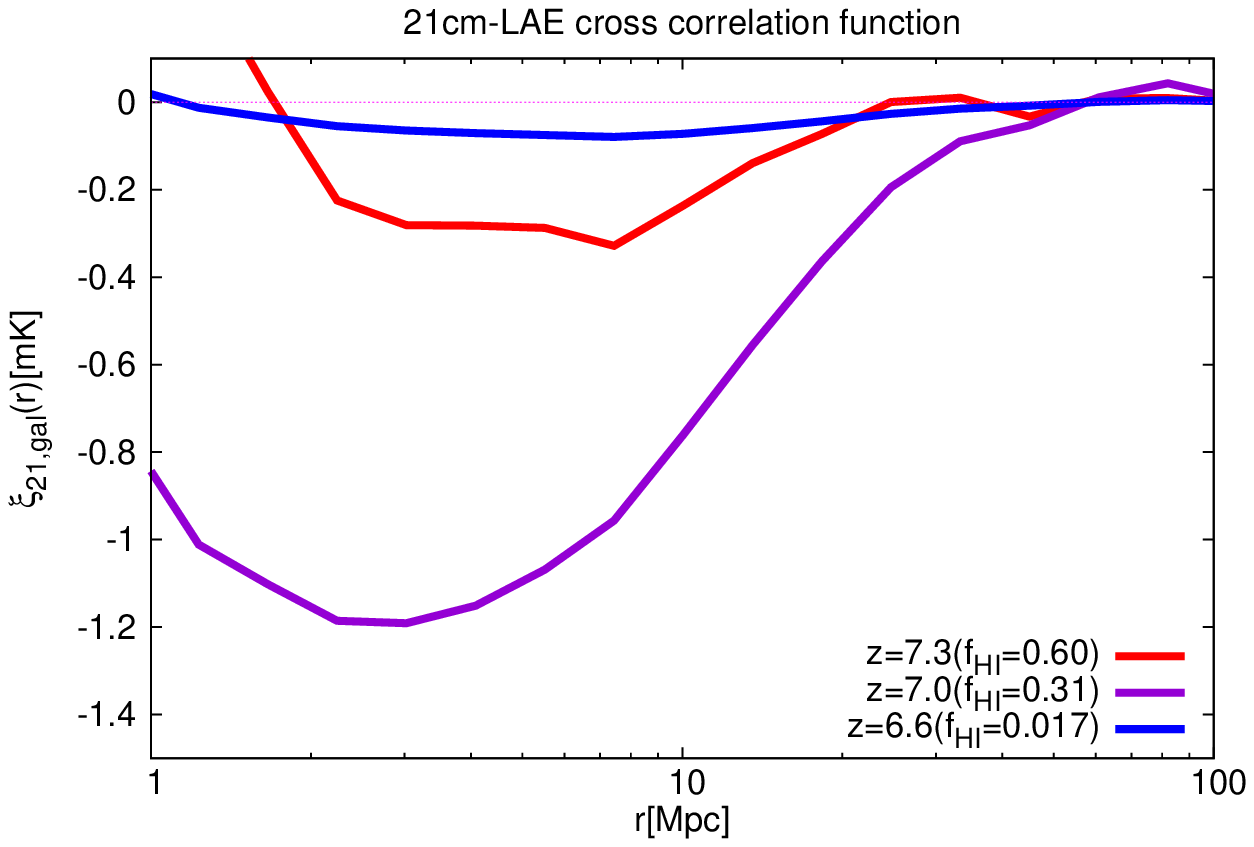}
\includegraphics[width=8cm,trim=0cm 0cm 0cm 0cm]{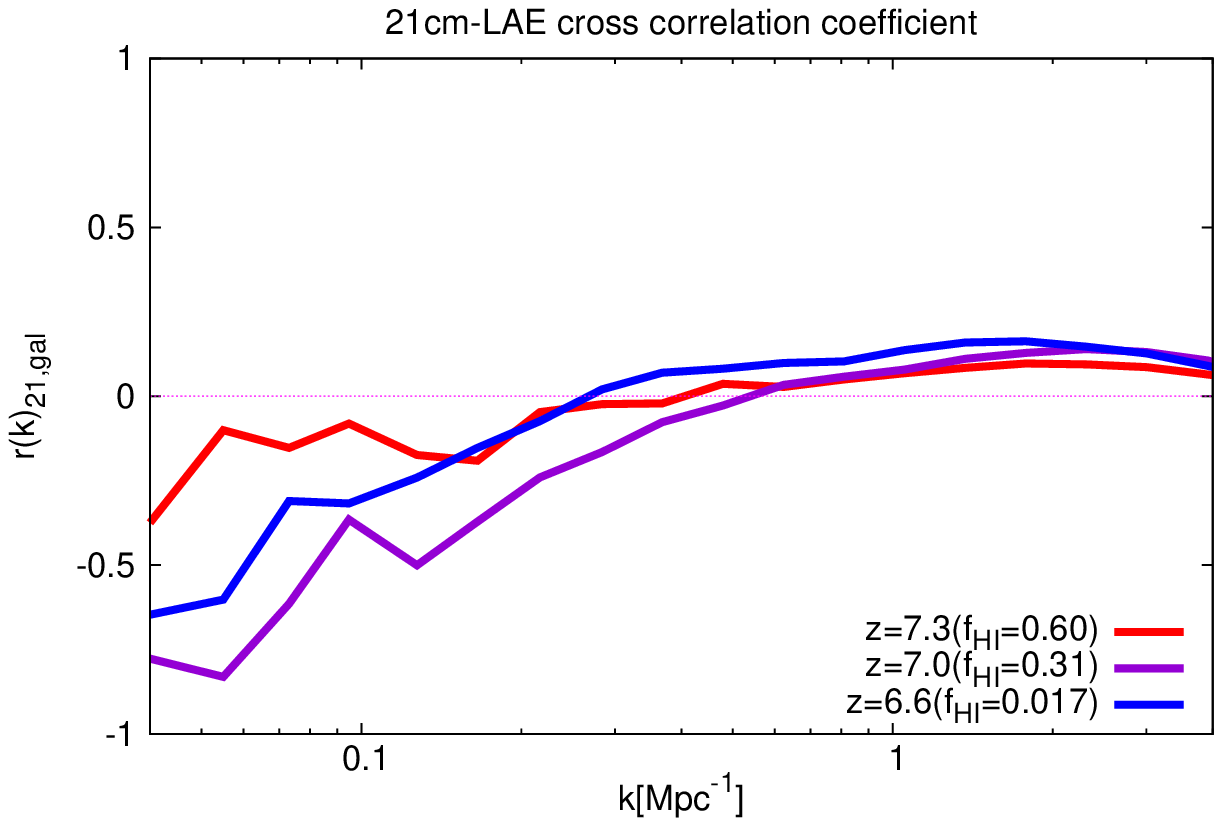}
\end{center}
\caption{21cm-LAE cross-power spectrum (top), cross-correlation function (center), and cross-correlation coefficient (bottom) in mid model. In the top figure the solid and the dotted line represent positive and negative value of the cross-power spectrum, respectively. We show results at (redshift $z$, neutral fraction $f_{\rm HI}$) = $(7.3, 0.60), (7.0, 0.31)$ and $(6.6, 0.017)$.}
\label{fig:cross_mid}
\end{figure}

First of all, we start by showing the redshift evolution of the 21cm-LAE cross-correlation statistics in our simulations. Fig.~\ref{fig:cross_mid} shows the 21cm-LAE cross-power spectrum, cross-correlation function, and cross-correlation coefficient at redshift $z=7.3$, $7.0$, $6.6$. Here, we counted LAEs which are brighter than the detectable luminosity in the Ultra-deep survey of Subaru HSC at redshift $z=6.6, 7.3$. We set the detectable luminosity at redshift $z=7.0$ by interpolating between $z=6.6$ and $z=7.3$. Generally, the cross-power spectrum has large absolute values when the average neutral fraction is close to 0.5, because the fluctuations in neutral fraction is maximum then. As to the sign, it has negative (positive) values at large (small) scales as seen in previous works. The positive correlation is considered to be caused by the correlation between the ionized region around the LAEs and the underdense region inside the ionized bubbles. The sign of the cross-power spectrum changes at $k \sim 0.3~{\rm Mpc}^{-1}$ at $z=6.6$ and $k \sim 0.8~{\rm Mpc}^{-1}$ at $z=7.3$. This scale is often called turnover scale and represents a typical size of ionized bubbles at a given epoch \citep{2009ApJ...690..252L}. These behaviors can also be seen in the cross-correlation coefficient (bottom of Fig.~\ref{fig:cross_mid}). While the negative correlation at large scales are relatively strong, the coefficient at small scales is positive but much smaller than unity so that the correlation is very weak.

The cross-power spectrum from our simulations has relatively large amplitudes at small scales compared to the previous works with semi-numerical methods\citep{2009ApJ...690..252L,2014MNRAS.438.2474P,2016MNRAS.459.2741S}. This is caused by the difference in the treatment of ionization state in high density regions. While the ionization fraction inside ionized bubbles is exactly equal to zero in most semi-numerical methods, because the recombination rate is properly taken into account in our simulations as described in Sec.3, high density regions inside ionized bubbles where LAEs often reside are slightly neutral in our calculation. These slightly neutral regions contribute to the cross-correlation and auto-correlation at small scales.

The cross-correlation function (center of Fig.~\ref{fig:cross_mid}) also shows the negative correlation at the associated scale. The cross-correlation function shows negative correlation at scales smaller than $\sim 40~{\rm Mpc}$ and has a large amplitude at $z = 7.0$. The negative correlation at small scales is caused by galaxy fluctuations embedded in mostly ionized regions. The amplitude of the negative correlation becomes larger from almost neutral state to half ionized state and it is largest when the half of the IGM is ionized\citep{2016arXiv160501734H}. The larger amplitude of $z=7.0$ at small scales indeed describes such the behavior.

Thus, we could confirm the qualitative features found in previous works with more realistic simulations with the improved treatment of the recombination rate and the clumping factor for the calculation of ionization structure.

\subsection{Detectability}

\begin{figure}
\begin{center}
\includegraphics[width=8cm,trim=0cm 0cm 0cm 0cm]{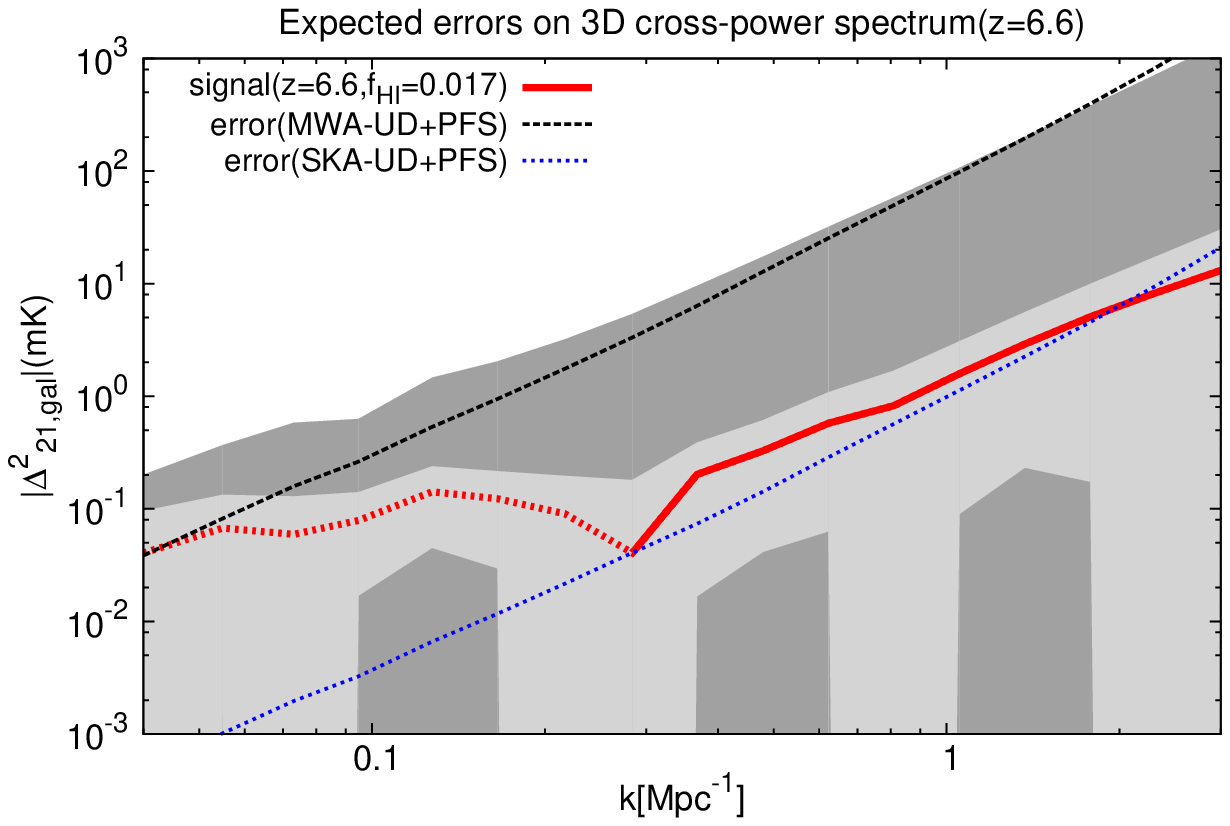}
\includegraphics[width=8cm,trim=0cm 0cm 0cm 0cm]{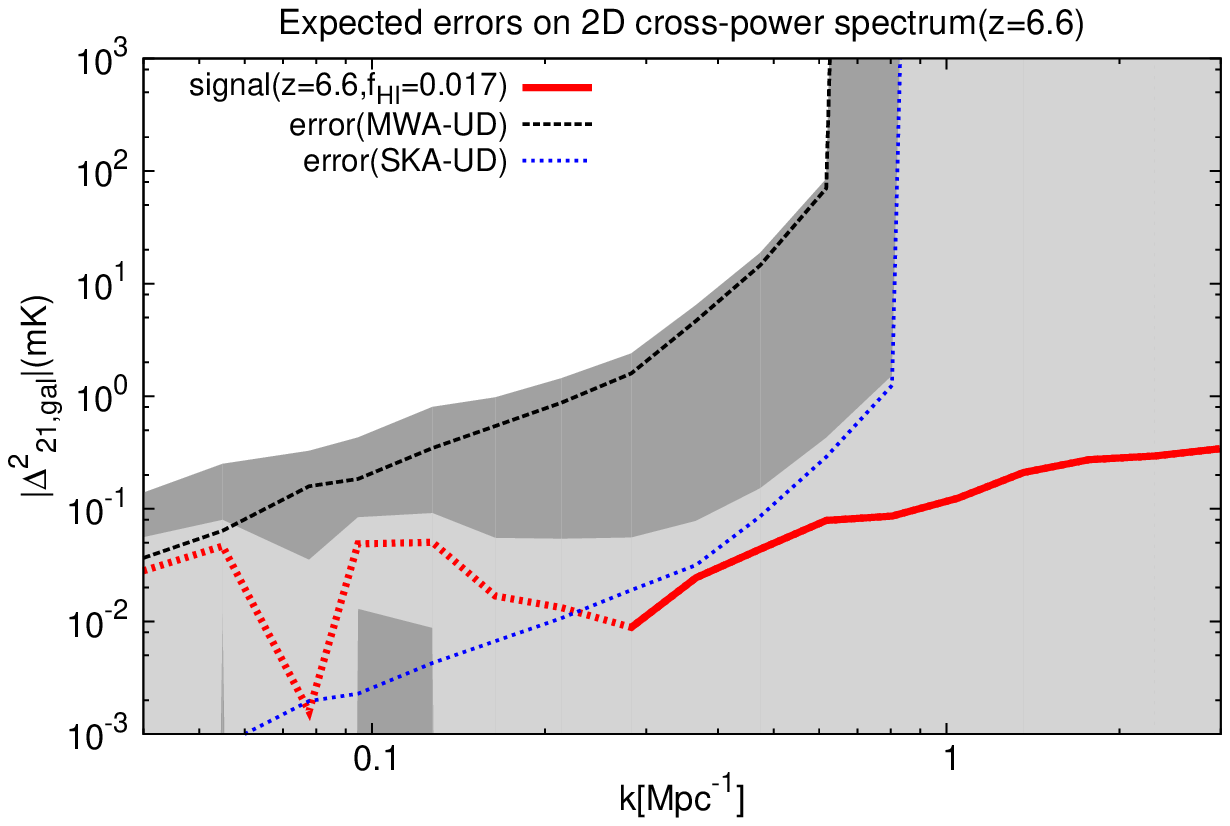}
\end{center}
\caption{MWA-Ultra deep and SKA-Ultra deep cross-correlation in the mid model at redshift $z=6.6$ with PFS (top) and without PFS (bottom). The red line shows the 21cm-LAE cross-power spectrum and the dark and light shadow show sample variance for the MWA and SKA, respectively. The dashed and dotted line show the sensitivity in the cross-correlation for the MWA and SKA, respectively. We note $k$ value on the 2D cross-power spectrum is computed by setting $k_{\parallel}=0$.}
\label{fig:ultra_mid_6.6}
\end{figure}

\begin{figure}
\begin{center}
\includegraphics[width=8cm,trim=0cm 0cm 0cm 0cm]{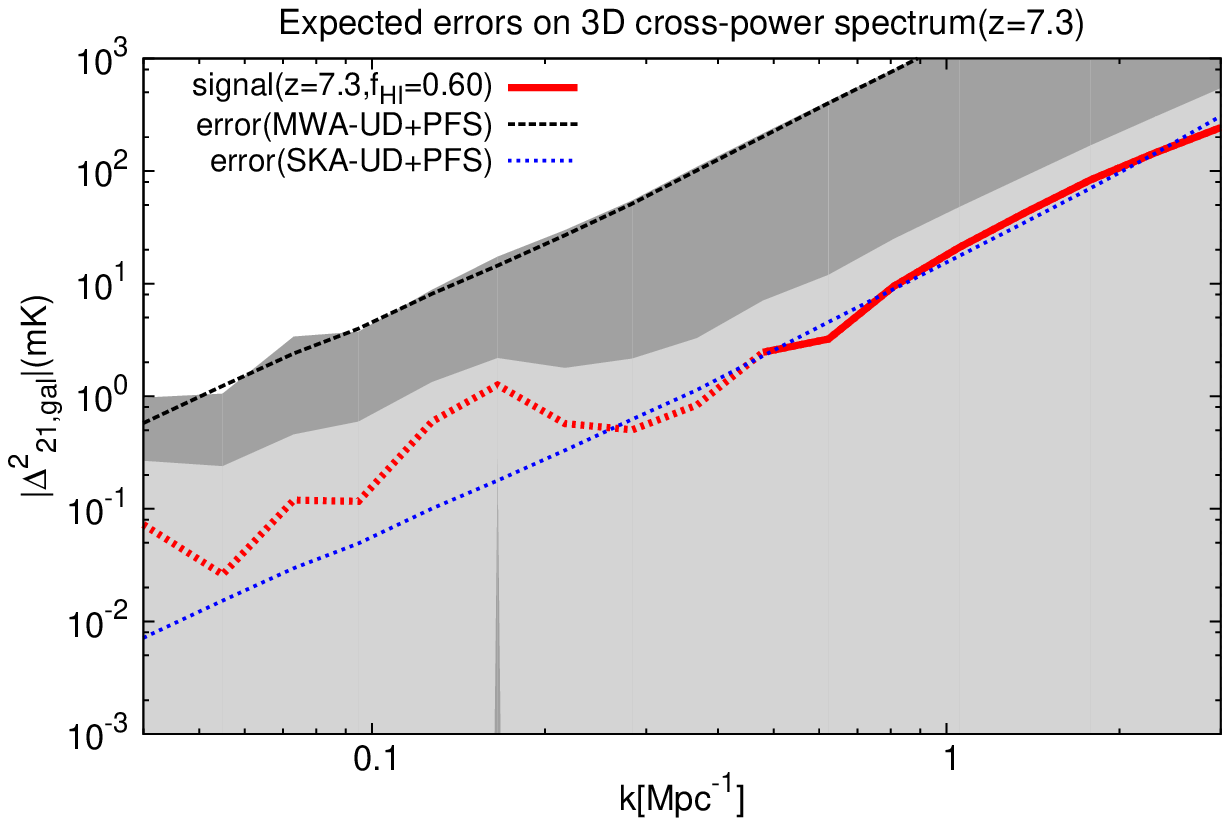}
\includegraphics[width=8cm,trim=0cm 0cm 0cm 0cm]{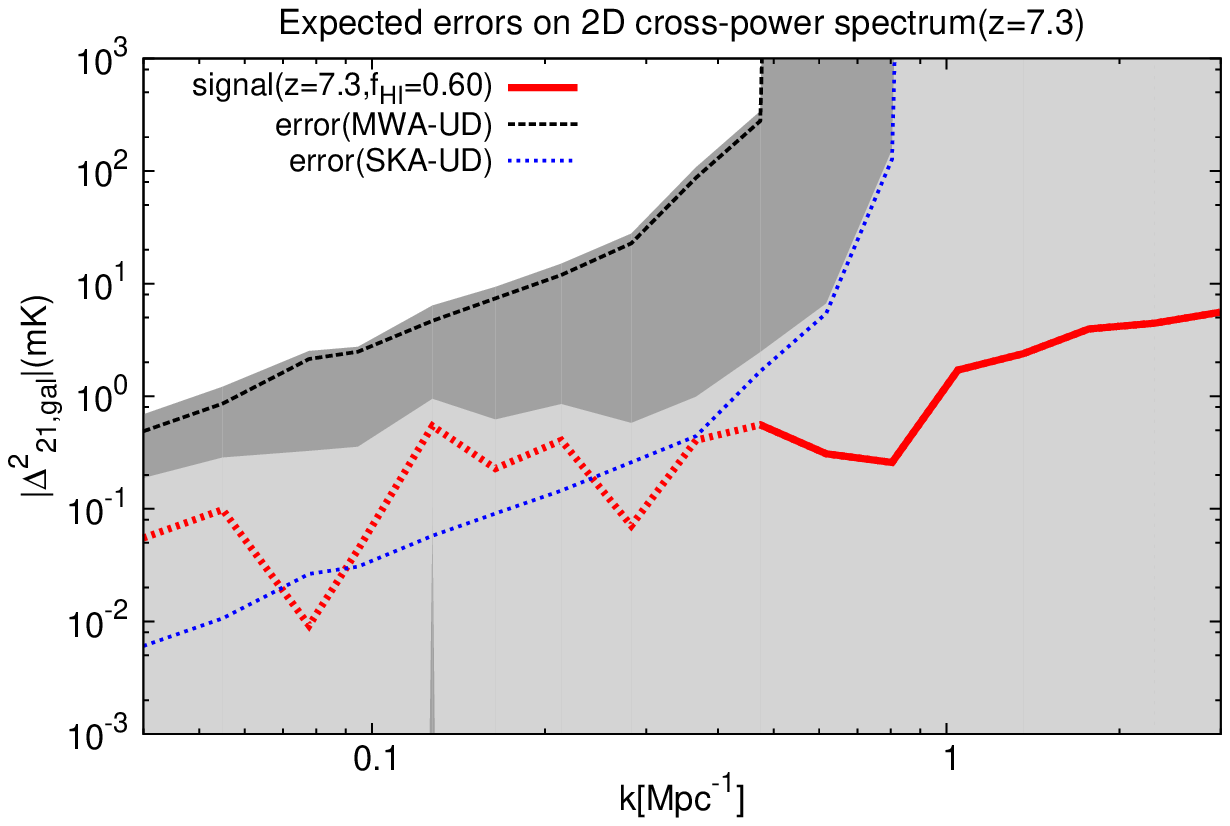}
\end{center}
\caption{Same as Fig.\ref{fig:ultra_mid_6.6}, but at redshift $z=7.3$.}
\label{fig:ultra_mid_7.3}
\end{figure}

\begin{figure}
\begin{center}
\includegraphics[width=8cm,trim=0cm 0cm 0cm 0cm]{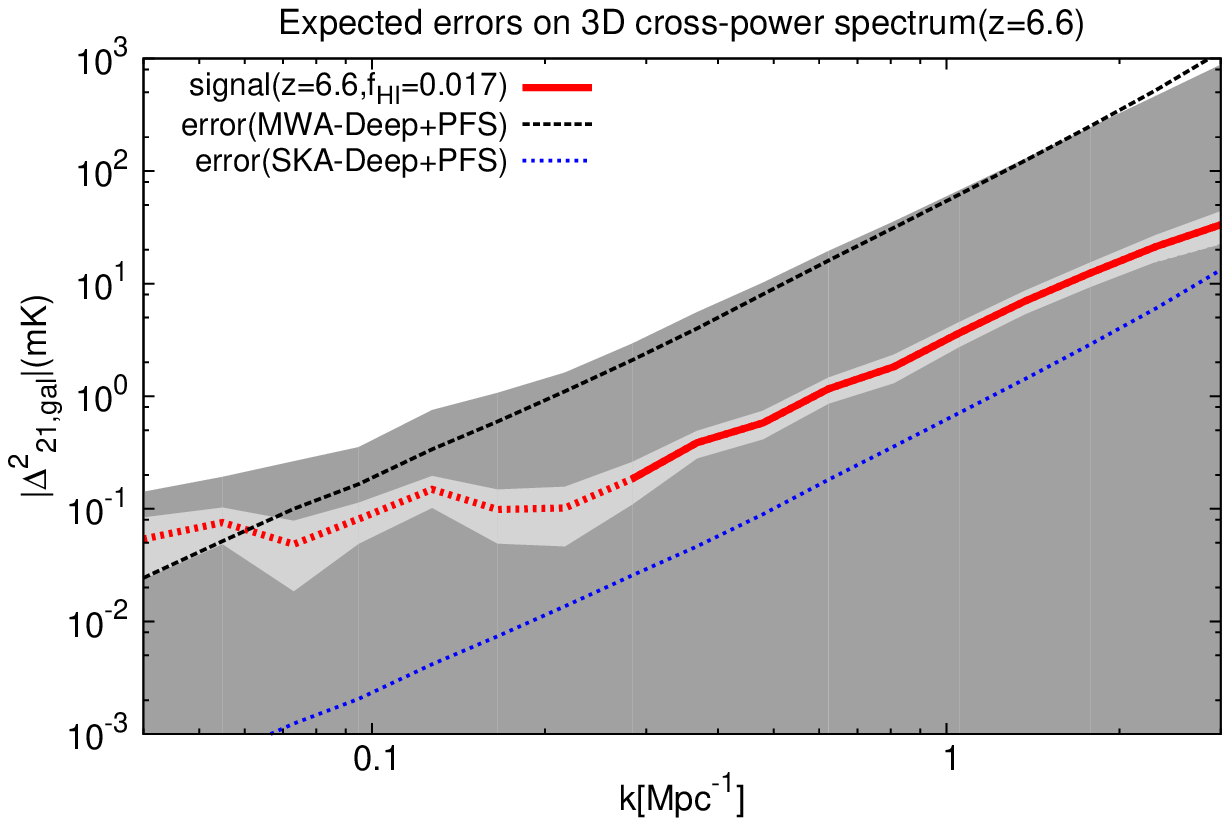}
\includegraphics[width=8cm,trim=0cm 0cm 0cm 0cm]{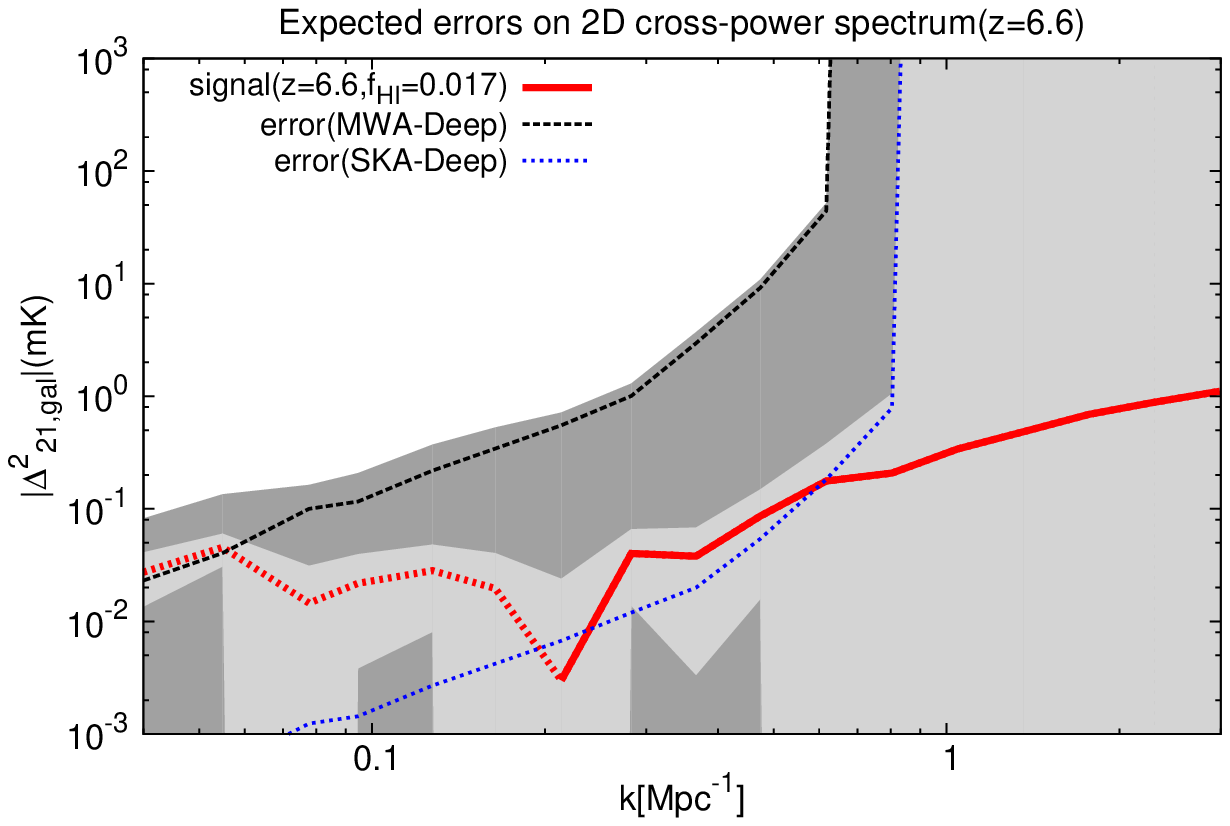}
\end{center}
\caption{MWA-Deep and SKA-Deep cross-correlation in the mid model at redshift $z=6.6$.}
\label{fig:deep_mid}
\end{figure}

In this subsection, we discuss the detectability of cross-correlation signal.
In Sec.5.1, we demonstrated the 3D cross-power spectrum signal in our simulation box. Here, we consider a case where PFS is unavailable and precise redshift information cannot be obtained, as well as a case where both HSC and PFS are available. As we mentioned before, in the former case, only 2D cross-power spectrum can be measured. To estimate the signal of this case, we integrate 21cm-line signal of a slice with the width of $40 \rm Mpc$, which corresponds to the redshift uncertainty of HSC, along the redshift direction. Actually, we generate 12 slices from our simulation box, changing the direction of integration, and adopt the median value of the signal. 
Figs.~\ref{fig:ultra_mid_6.6}, \ref{fig:ultra_mid_7.3}, and \ref{fig:deep_mid} show the 3D and 2D cross-power spectra, for Ultra-deep surveys at redshifts $z=6.6$ and $z=7.3$, and Deep survey, respectively.
Comparing with the 3D signal (top panels) with PFS, the 2D signal (bottom panels) reduces at small scales because fluctuations is smoothed by the integration and information on $k_{\parallel}$ modes is lost.

In these figures, sample variance and the sensitivities for MWA and SKA are also shown. First, let us discuss the detectability for MWA with Ultra-deep survey (Figs.~\ref{fig:ultra_mid_6.6} and \ref{fig:ultra_mid_7.3}). The sensitivity is better for the case with PFS compared with the case without PFS, especially at small scales. At $z = 6.6$, the sensitivity is comparable to the average signal amplitude at large scales ($k \lesssim 0.1~{\rm Mpc}^{-1}$). However, due to the large sample variance, the signal may not be detectable in sky areas with smaller signal amplitudes than the average. The situation is much worse at $z = 7.3$ where the sensitivity is worse than the average signal at least by one order of magnitude. The total S/N ratio, considering sample variance as well as observational uncertainties, is 0.42 (0.38) with (without) PFS at $z=6.6$ and 0.13 (0.11) with (without) PFS at $z=7.3$ (see Table \ref{tb:SN}). PFS does not increase the total S/N ratio so much, because MWA is not sensitive at small scales. Thus, it is difficult to detect signal with a combination of MWA and Ultra-deep survey even with a follow-up of PFS.

Next, let us discuss the detectability for MWA with Deep survey at $z=6.6$ (Fig.~\ref{fig:deep_mid}). The qualitative features of the sensitivities and sample variance are very similar to the case with Ultra-deep survey. However, the wider survey area compared to Ultra-deep survey reduces the sample variance and compensates for the larger detection limit of LAEs which leads to a smaller LAE density and then a larger shot noise. Consequently, the total S/N ratio is slightly better, 1.0 and 0.73 with and without PFS, respectively. Thus, it will be possible to detect the signal if the survey area has larger cross-correlation amplitude than the average.

\begin{table}
\begin{center}
\caption{Total S/N ratio of the cross-power spectrum for the mid model. In Deep field survey, the S/N ratios are also shown in the case of extended survey area and observation time per pointing (depth) by a factor of 3, respectively.}
\renewcommand{\arraystretch}{1.3}
\begin{tabular}{|c|c|c|c|c|c|c|} \hline
 &  PFS  & $z$ & UD & Deep  & area $\times 3$  & depth $\times 3$ \\ \hline \hline
            &\multirow{2}{*}{on} & 6.6   & 0.42        & 1.0     &1.7  & 1.2   \\ \cline{3-7}
\multirow{2}{*}{MWA}   &     & 7.3   & 0.13       & -     & -  & -  \\ \cline{2-7}
           &\multirow{2}{*}{off} & 6.6   & 0.38        & 0.73     &1.3  & 1.1   \\ \cline{3-7}
           &      & 7.3   & 0.13       & -     & -  & -  \\[-3pt] \hline
           &\multirow{2}{*}{on} & 6.6   & 4.1         & 11    &20  & 11        \\ \cline{3-7}
\multirow{2}{*}{SKA}    &      & 7.3   & 2.6         & -   & -  & -    \\ \cline{2-7}
           &\multirow{2}{*}{off} & 6.6   & 2.8         & 5.1    & 8.9  & 8.4        \\ \cline{3-7}
           &      & 7.3   & 1.9         & -   & -  & -    \\[-3pt] \hline
\end{tabular}
\label{tb:SN}
\end{center}
\end{table}

The situation changes drastically with the SKA1-low. Thanks to the large effective area, the sensitivities improve drastically and much larger S/N ratio is expected (see Table~\ref{tb:SN}). The signal could be detected even at $z = 7.3$ with PFS (S/N$\sim$3), while it is marginal without PFS. Thus, with the SKA, we could study the evolution of cross-correlation at the late stage of the EoR. Further, in case of $z = 6.6$, the SKA can probe much smaller scales than the MWA. Especially, with PFS, it will be possible to detect the signal at scales as small as $k \sim 1~{\rm Mpc}^{-1}$ and the turnover of the cross correlation could be detected. This point will be discussed again in the next section. We note that some of the reduced sample variance exhibited at $k \sim 0.13~{\rm Mpc}^{-1}$, $0.5~{\rm Mpc}^{-1}$, and $1.3~{\rm Mpc}^{-1}$ in top panel of Fig.\ref{fig:ultra_mid_6.6} is caused by the slight change of the signal since the sample variance is comparable to the signal at all scales. For example, the sample variance seems to be small at $k \sim 0.5~{\rm Mpc}^{-1}$ and $1.3~{\rm Mpc}^{-1}$ because the signal is coincidentally slightly smaller than the sample variance at $k \sim 0.8~{\rm Mpc}^{-1}$.

\begin{figure}
\begin{center}
\includegraphics[width=8cm,trim=0cm 0cm 0cm 0cm]{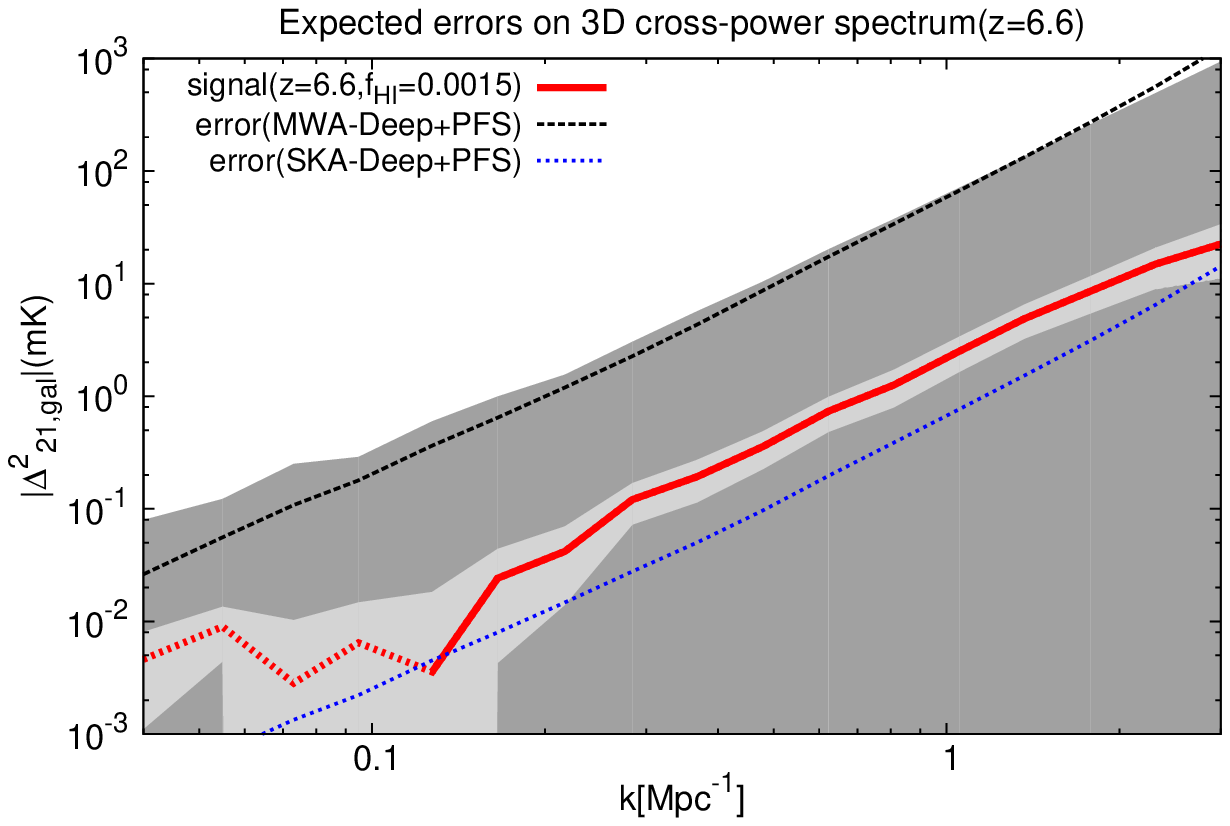}
\includegraphics[width=8cm,trim=0cm 0cm 0cm 0cm]{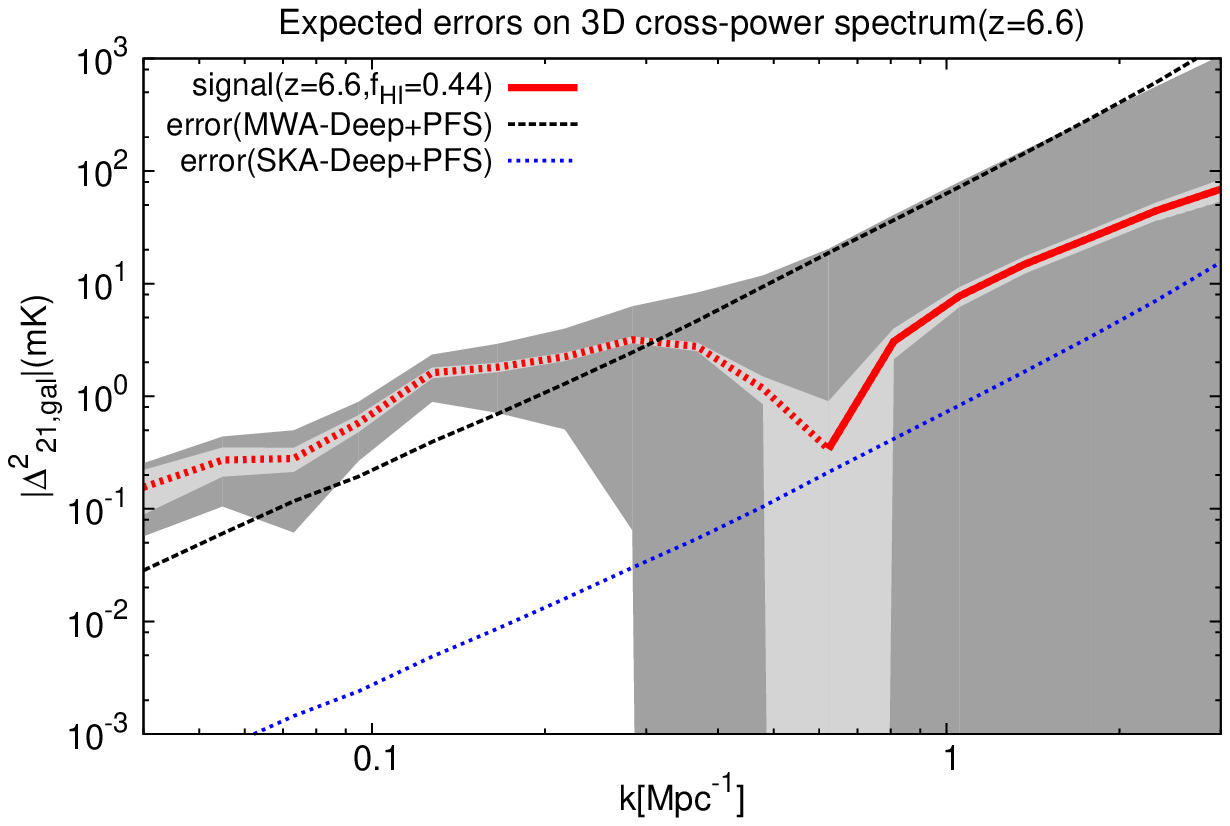}
\end{center}
\caption{Same as Fig.~\ref{fig:deep_mid} (Deep survey with PFS at $z=6.6$), but in the early model (top) and late model (bottom).}
\label{fig:deep_early_late}
\end{figure}

Next, let us compare the detectability for the three EoR models. Fig.~\ref{fig:deep_early_late} represents the signal and sensitivities for the early and the late models with MWA-Deep survey and SKA-Deep survey with PFS, respectively. The average neutral fraction at $z=6.6$ is 0.0015 and 0.44 for the early and the late models, respectively, while it is 0.017 for the mid model. The amplitude of the cross-correlation signal is largely determined by the average neutral fraction, and the signal is smaller (larger) for the early (late) model compared with the mid model. The ratios of the signal amplitude at large scales are about 3 between the early and the mid models and between the mid and the late models. As we can see, the detectability strongly depends on the EoR model. For the early model, it is very hard for the MWA to detect the signal even if PFS is available. The S/N ratios are 0.14 and 0.083 with and without PFS, respectively, while they are still relatively high for the SKA: 7.5 and 5.1 with and without PFS, respectively. On the other hand, for the late model, the MWA could detect the signal even without PFS, while the signal could be detected at relatively small scales ($k \lesssim 0.3~{\rm Mpc}^{-1}$) with PFS.

\begin{table}
\begin{center}
\caption{Comparison of total S/N ratio of the cross-power spectrum in the early, mid, and late models. The S/N ratios are shown in the cross-correlation with Deep field survey.}
\renewcommand{\arraystretch}{1.3}
\begin{tabular}{|c|c|c|c|c|} \hline
 &  PFS   & early & mid  & late \\ \hline \hline
 \multirow{2}{*}{MWA}  & on  & 0.14        & 1.0     &4.3    \\ \cline{2-5}  
           & off  & 0.083        & 0.73     &4.3  \\ \cline{1-5}
\multirow{2}{*}{SKA}  & on  & 7.5         & 11    &31       \\ \cline{2-5} 
           & off  & 5.1         & 5.1    & 20   \\[-3pt] \hline
\end{tabular}
\label{tb:SN_model}
\end{center}
\end{table}

\begin{figure}
\begin{center}
\includegraphics[width=8cm,trim=0cm 0cm 0cm 0cm]{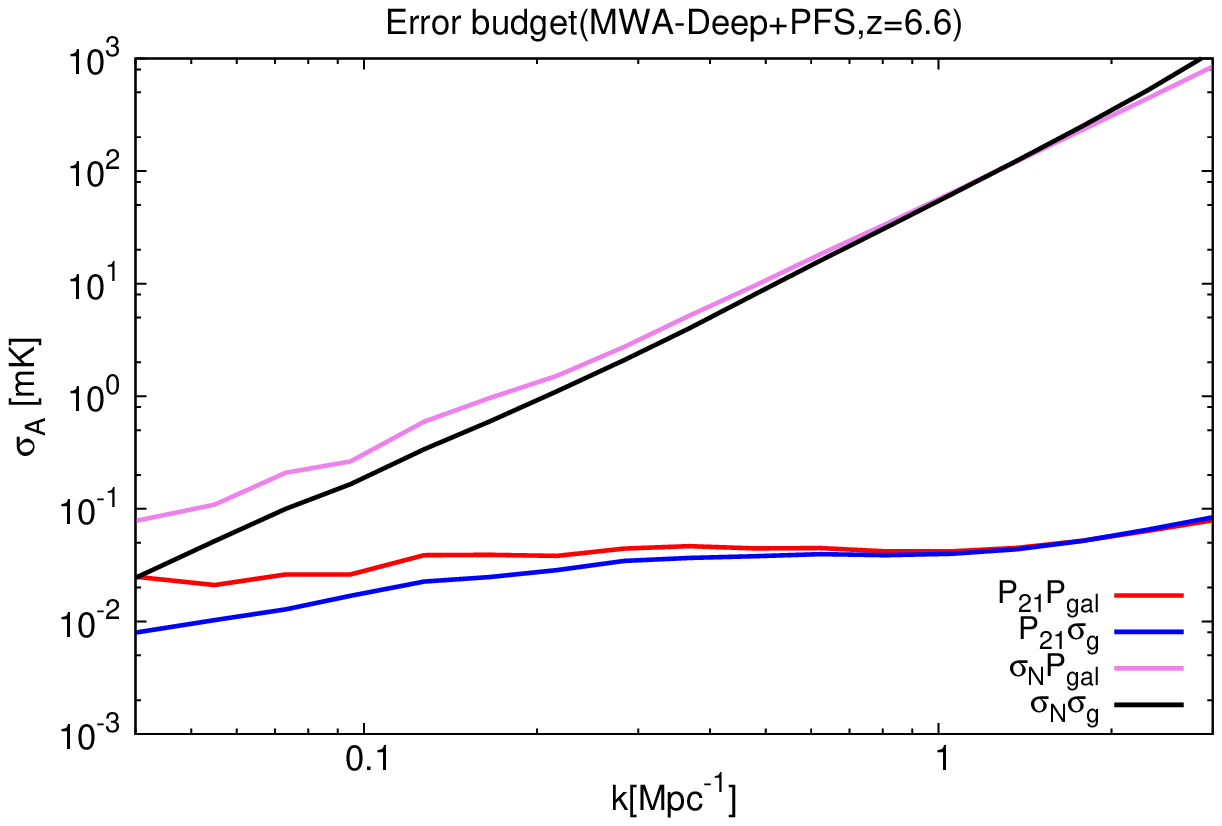}
\includegraphics[width=8cm,trim=0cm 0cm 0cm 0cm]{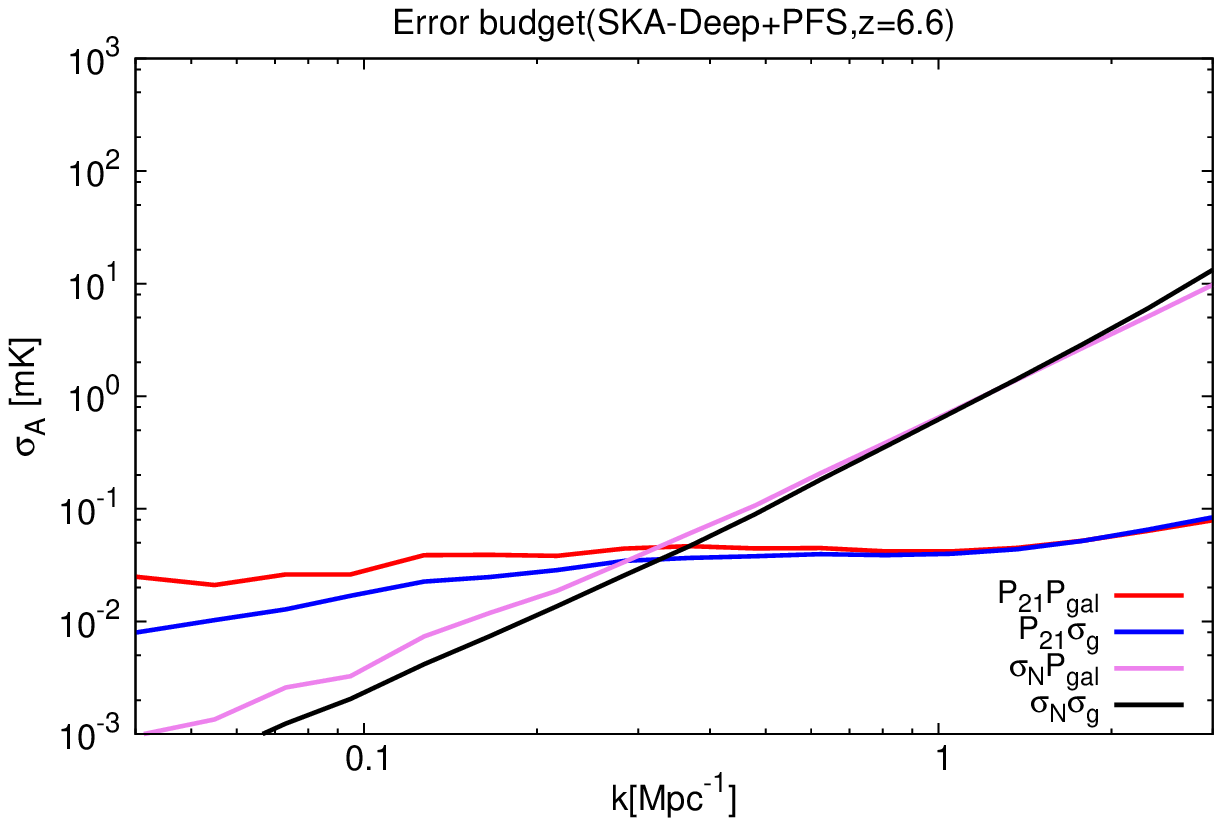}
\end{center}
\caption{Error budgets of the sensitivity for MWA-Deep survey (top) and SKA1-Deep survey (bottom) with PFS for the mid model. The red, blue, pink, black lines show the components of $\sigma_{\rm A}$ as in Eq.~(\ref{eq:budget}); $P_{21}P_{\rm gal}$, $P_{21}\sigma_{\rm g}$, $\sigma_{\rm N}P_{\rm gal}$, and $\sigma_{\rm N}\sigma_{\rm g}$, respectively.}
\label{fig:error budget}
\end{figure}

\begin{figure}
\begin{center}
\includegraphics[width=8cm,trim=0cm 0cm 0cm 0cm]{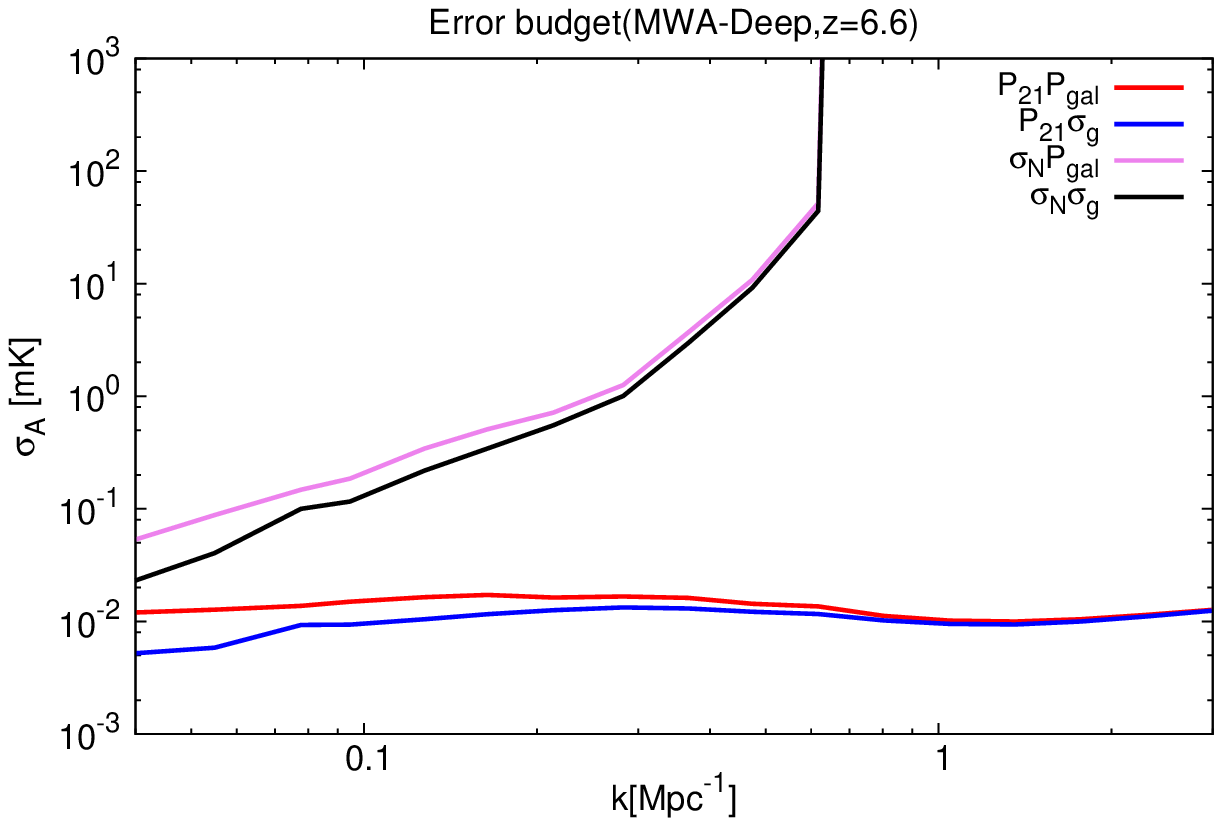}
\includegraphics[width=8cm,trim=0cm 0cm 0cm 0cm]{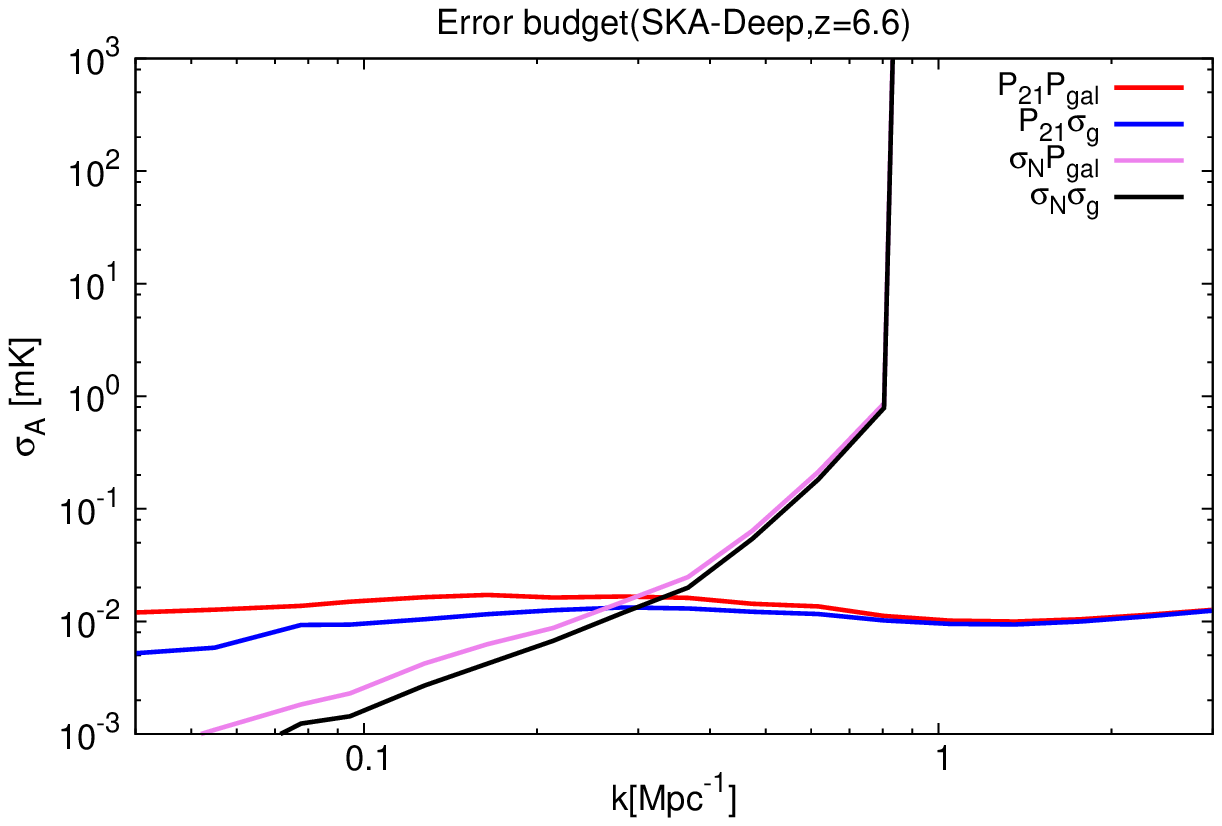}
\end{center}
\caption{Same as Fig.\ref{fig:error budget}, but without PFS}
\label{fig:error budget_2D}
\end{figure}

Next, to understand the sensitivity curves given above, we compare error components in Eq.~(\ref{eq:budget}): $P_{21}P_{\rm gal}$, $P_{21}\sigma_{\rm g}$, $\sigma_{\rm N}P_{\rm gal}$ and $\sigma_{\rm N}\sigma_{\rm g}$. The first one is a pure sample variance, the second and third ones are combinations of sample variance and observation errors, and the last one is a pure observational error. We do not show $P^2_{21,\rm gal}$ because it is always smaller than $P_{21}P_{\rm gal}$ by a factor of the correlation coefficient. Fig.~\ref{fig:error budget} shows the error budgets of MWA-Deep survey and SKA-Deep survey with PFS for the mid model, where the number of {\bf k} modes in Eq.\ref{eq:error-volume} is taken into account for each components. For MWA-Deep survey, $\sigma_{\rm N}P_{\rm gal}$ and $\sigma_{\rm N}\sigma_{\rm g}$ are dominant at all scales. Therefore, a reduction in $\sigma_{\rm N}$, the thermal noise of the MWA, by increasing observing time and/or number of antennas is effective to enhance the detectability. On the other hand, in case of the SKA, $P_{21}P_{\rm gal}$ and $P_{21}\sigma_{\rm g}$, that is, the sample variance terms of 21cm-line are dominant at large scales so that widening the survey area is the best way to increase the S/N ratio. Since the thermal-noise terms are sub-dominant, reducing the observing time, e.g., $\sim 100$ hours, does not affect the detectability at large scales significantly. Observing time as short as 10 hrs will be enough if we focus on the largest scales of $k \lesssim 0.1~{\rm Mpc}^{-1}$. Contrastingly, at small scales, the sensitivity is limited by $\sigma_{\rm N}P_{\rm gal}$ and $\sigma_{\rm N}\sigma_{\rm g}$ as the MWA case.
Fig.~\ref{fig:error budget_2D} shows the error budget without PFS. Although $\sigma_{\rm N}P_{\rm gal}$ and $\sigma_{\rm N}\sigma_{\rm g}$ are enhanced at small scales, the dominant error components for both of the MWA and SKA are the same as the case with PFS. Therefore, the best way to enhance the detectability is still the same even if PFS is unavailable.

To develop strategy for increasing S/N ratio, we consider two extensions of HSC Deep survey with (1) a larger survey area and (2) a longer observation time per pointing, by a factor of 3, respectively. Note that these two options need the same amount of extra observation time. In increasing the survey area, the area of 21cm-line observation should also be widened. However, because the MWA and SKA-low have much larger field-of-view than HSC, we assume the survey area of 21cm-line observations is always larger than that of LAE survey. Further, we assume that the detection limit of LAEs is inversely proportional to the square root of observation time per pointing. 

\begin{figure}
\begin{center}
\includegraphics[width=8cm,trim=0cm 0cm 0cm 0cm]{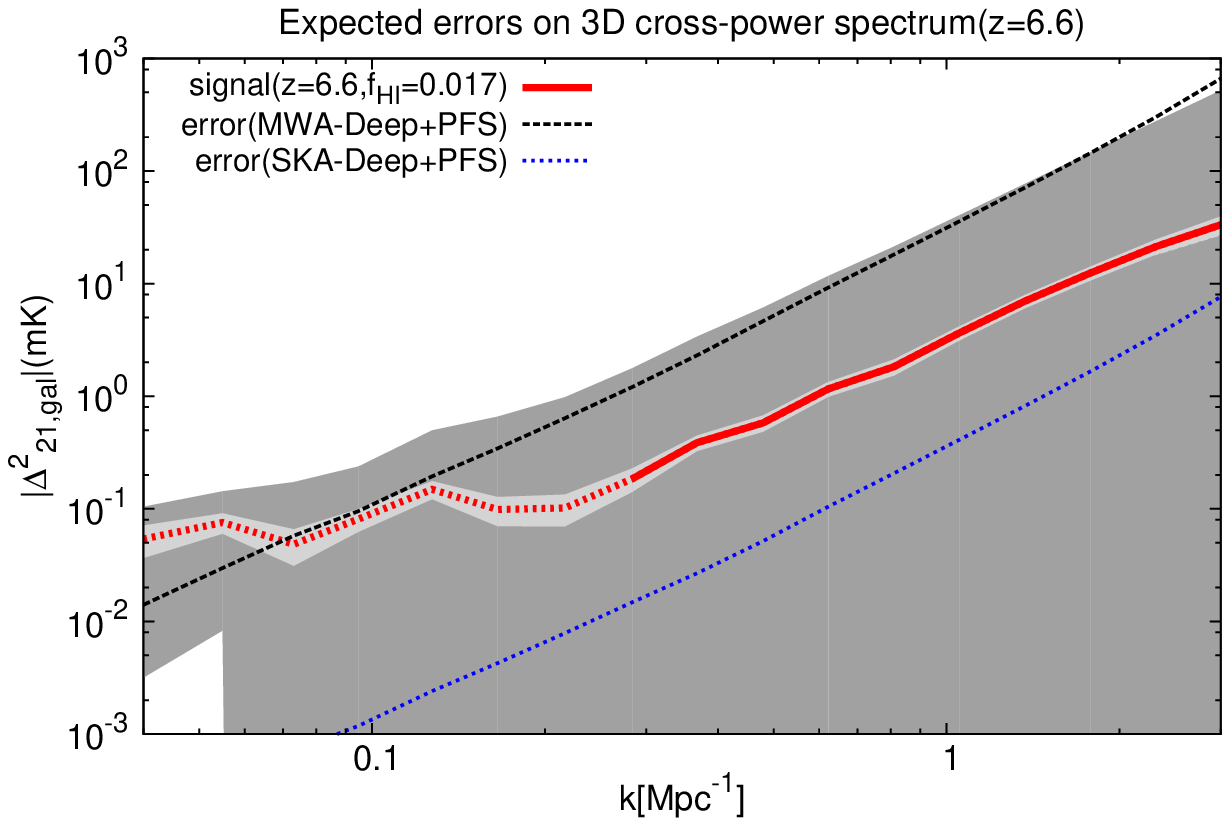}
\includegraphics[width=8cm,trim=0cm 0cm 0cm 0cm]{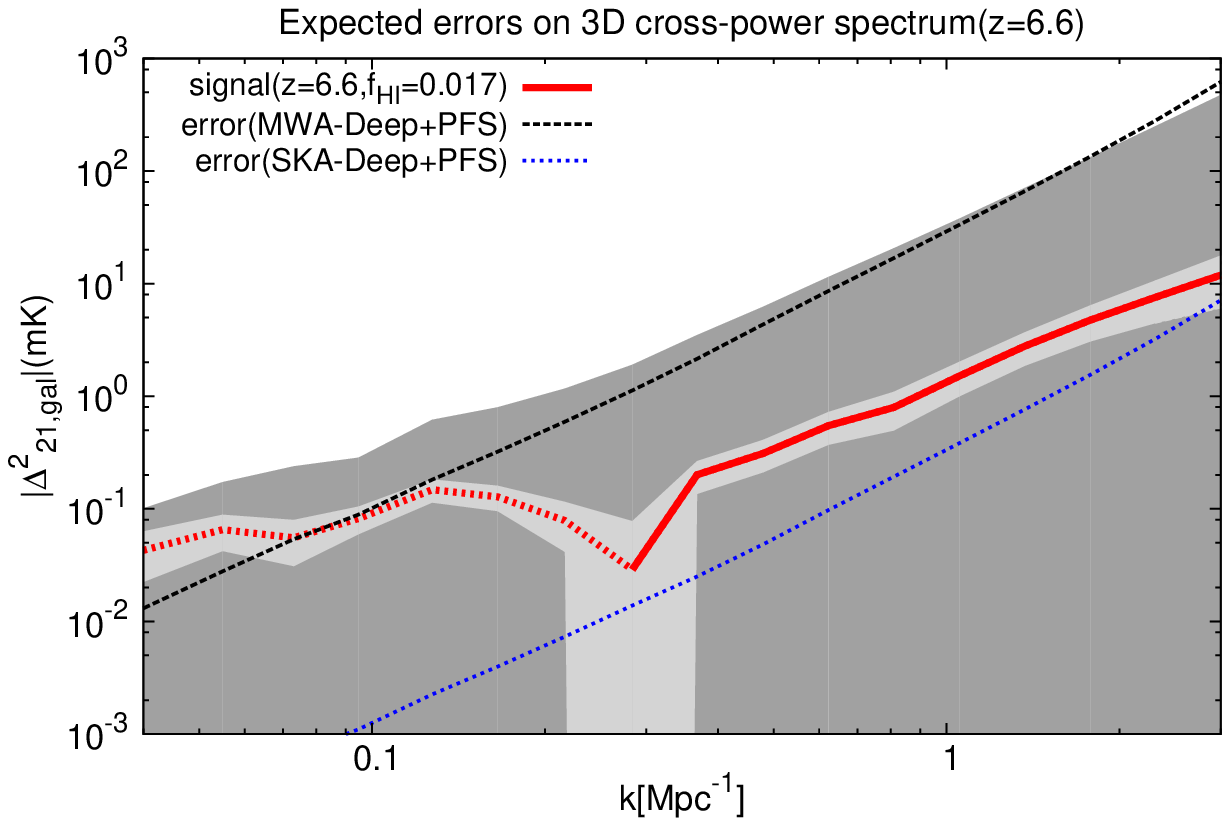}
\end{center}
\caption{Same as Fig.\ref{fig:deep_mid} (Deep survey with PFS, mid model), but with extended survey area (top) and observation time per pointing (bottom) by a factor of 3.}
\label{fig:deep_3}
\end{figure}

Fig.~\ref{fig:deep_3} shows the results of the two options for the MWA and SKA, respectively. In Table~\ref{tb:SN}, the total S/N ratios for these cases are shown. The S/N ratio is generally improved but the option (1) is significantly more effective. This is because, as shown in Eq.~(\ref{eq:error-volume}), the increase in the survey area (and then the survey volume) reduces both observational errors ($\sigma_{\rm N}$ and $\sigma_{\rm g}$) and sample variances ($P_{21}$ and $P_{\rm gal}$), while the survey depth is related to only the shot noise of galaxies ($\sigma_{\rm g}$). Another reason is that, as we saw in Fig.~\ref{fig:error budget}, the error components including $P_{\rm gal}$ are always larger than those including $\sigma_{\rm g}$. Here, it should be noted that the effect of changing the survey depth depends on the LAE luminosity function at the faint end ($L_{\alpha} \sim 10^{42.5}~{\rm erg/s}$, see Fig.~\ref{fig:LF}). In case of a steeper luminosity function (fixing the bright end), a deeper survey results in more LAE density and smaller shot noise.

\begin{figure}
\begin{center}
\includegraphics[width=10cm,trim=1cm 0cm 1cm 0cm]{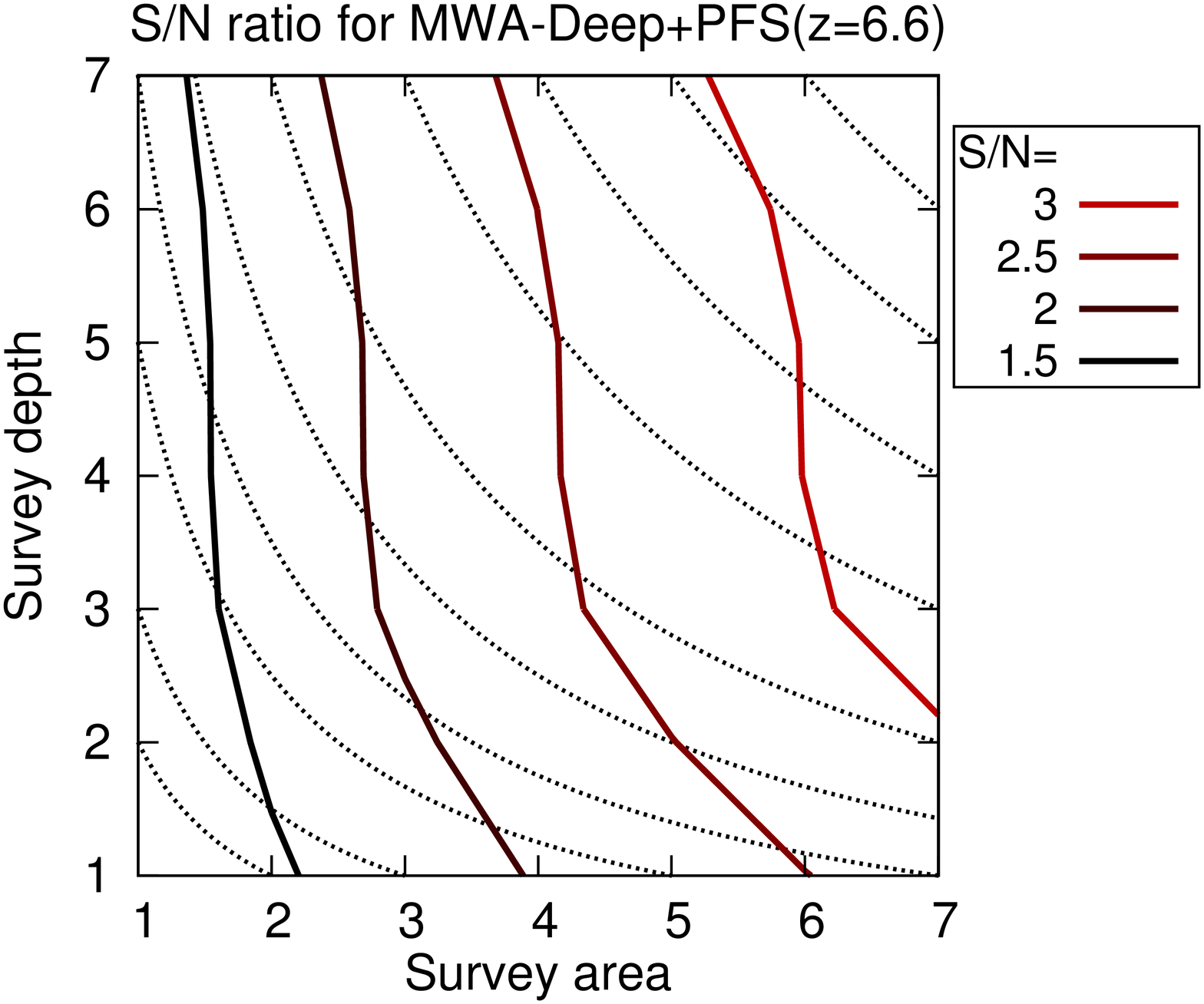}
\includegraphics[width=10cm,trim=1cm 0cm 1cm 0cm]{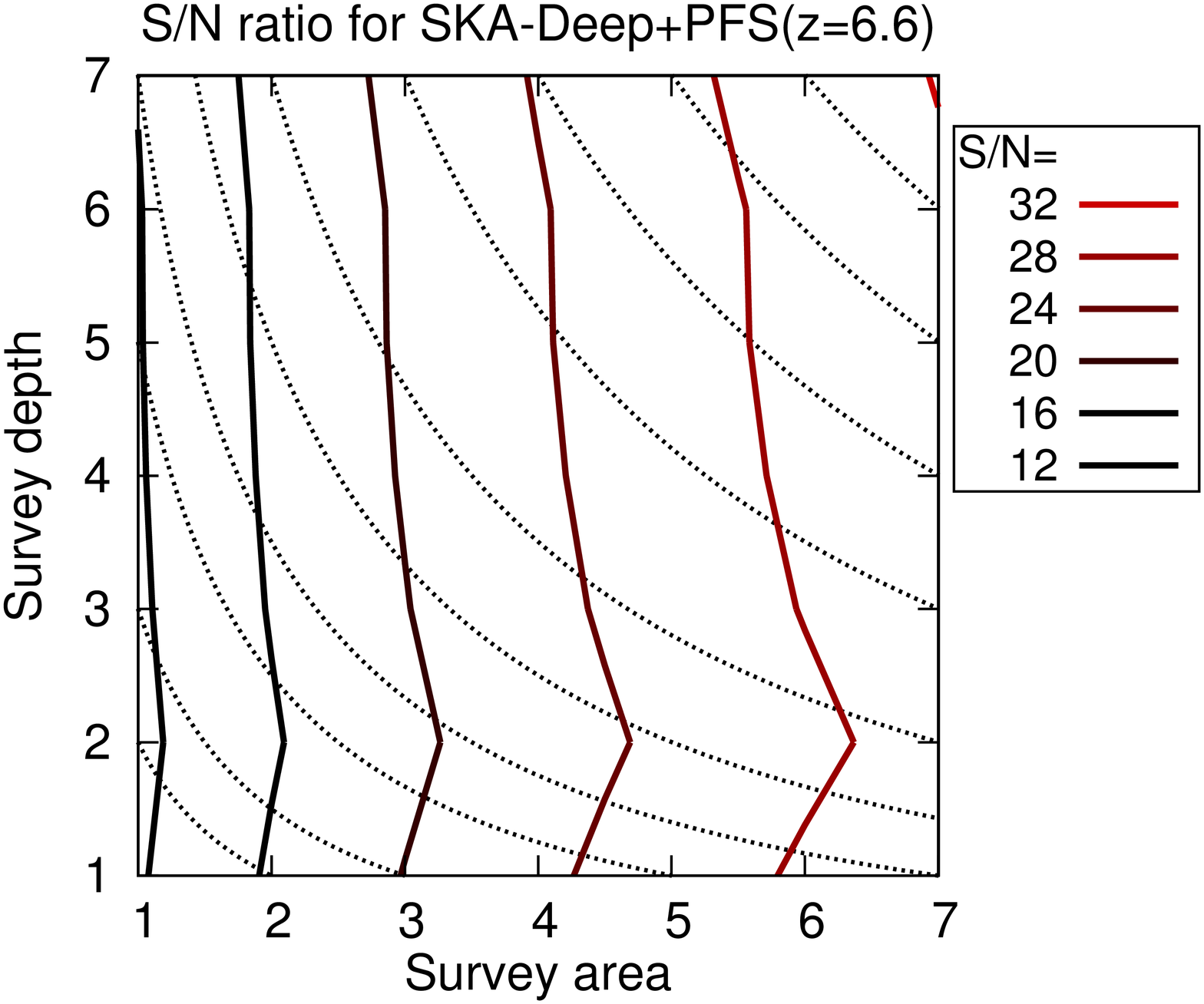}
\end{center}
\caption{S/N ratio contour in HSC survey area-depth plane for MWA (top) and SKA (bottom) with PFS follow-up observations. The area and depth are normalized by their fiducial values. The solid lines represent the S/N contour lines and the dotted lines represent equal survey-time lines. Mid EoR model is used.}
\label{fig:SN_mid_contour}
\end{figure}

\begin{figure}
\begin{center}
\includegraphics[width=10cm,trim=1cm 0cm 1cm 0cm]{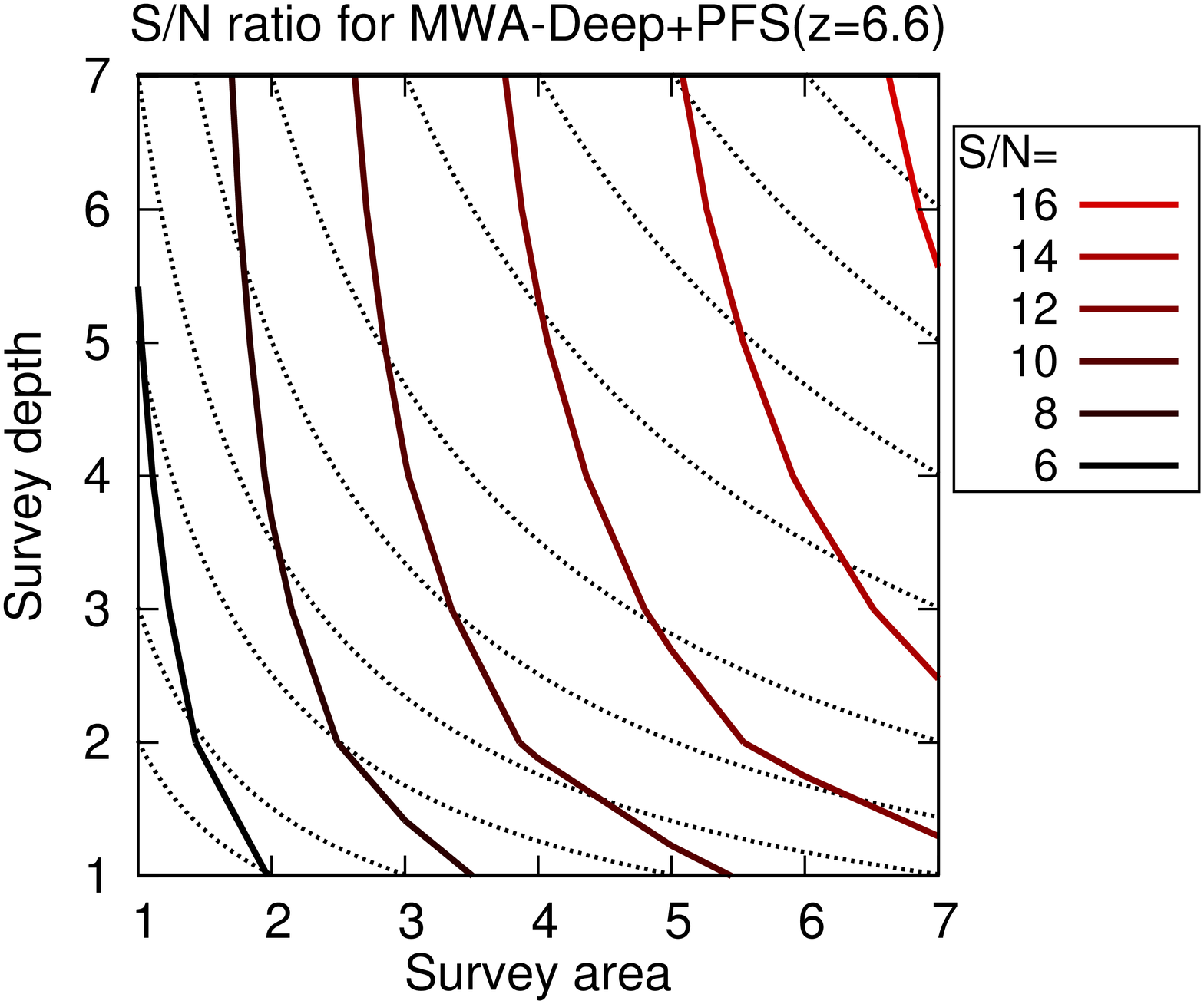}
\includegraphics[width=10cm,trim=1cm 0cm 1cm 0cm]{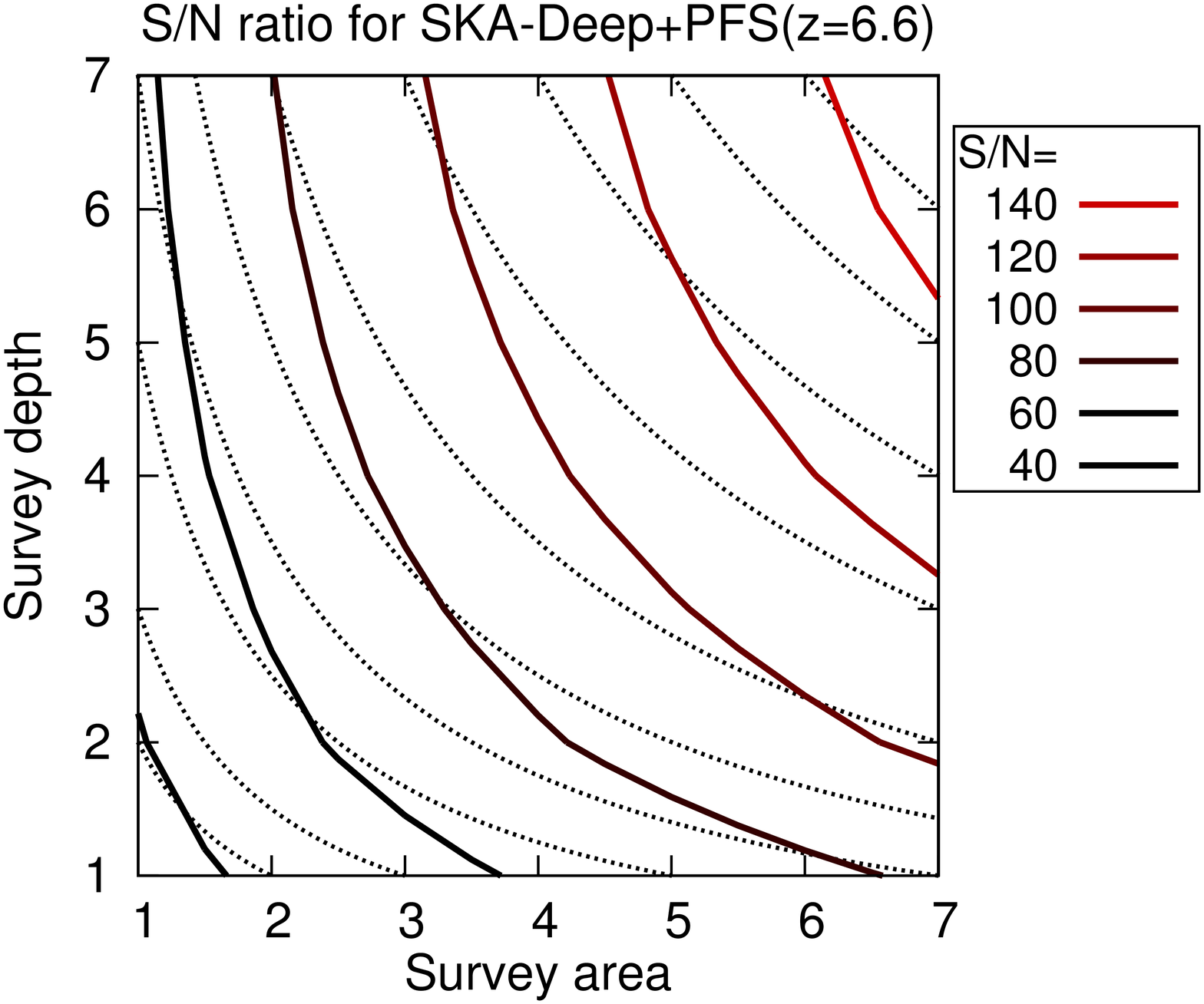}
\end{center}
\caption{Same as Fig.\ref{fig:SN_mid_contour}, but for the late model.}
\vspace{5mm}
\label{fig:SN_late_contour}
\end{figure}

More general results are shown in Figs.~\ref{fig:SN_mid_contour} and \ref{fig:SN_late_contour} which represent the contours of the total S/N ratio of MWA-Deep survey and SKA-Deep survey for the mid and late models in survey area-depth plane. We can see that, for a fixed survey area, deeper observation of LAEs does not improve the S/N ratio so much. Thus a wide and shallow LAE survey leads to a larger S/N ratio given a fixed total survey time. This is an important implication for future observation strategy.

The behavior of the contours is not simple for the case of SKA with mid model (bottom panel of Fig.~\ref{fig:SN_mid_contour}). Actually, the signal decreases at small scales (k > 0.3 /Mpc) as the LAE survey gets deeper while the signal is unchanged at large scales (k < 0.3 /Mpc). This is because the fluctuations generated by small scale clustering of the brighter LAEs are smoothed by the fainter LAEs. On the other hand, the observation errors are reduced for a deeper LAE survey. Altogether, S/N increases (decreases) at large (small) scales as a function of the survey depth. As we can see in Figs. \ref{fig:deep_mid} and \ref{fig:deep_early_late}, in case of MWA with mid and late models and SKA with late model, the total S/N is contributed mostly from large-scale observation so that the total S/N increases monotonically as a function of LAE survey depth. Contrastingly, in case of SKA with mid model, the contributions from large and small scales are comparable. This is why the the behavior of the total S/N is not monotonic.

\section{Summary and Discussion}

In this paper, we investigated the detectability of the 21cm-LAE cross-correlation signal, which is potentially powerful to reduce foreground contamination of 21cm-line signal. The 21cm-line emission is expected to have negative correlation with LAE clustering if LAEs are main sources of ionizing photons. First, we have confirmed the qualitative features of the signal discovered in previous works under our realistic numerical simulations. The cross-power spectrum is negative at large scales and positive at small scales, although the correlation coefficient is close to zero at small scales.

Next, we computed the sensitivity, assuming combinations of 21cm-line observations by the MWA and SKA and LAE surveys by HSC with/without a follow-up observation by PFS, and compared with the signal. At $z = 6.6$, the S/N ratio is always better for Deep survey compared with Ultra-Deep survey, while only the latter can probe $z = 7.3$. We found that the detectability strongly depends on the EoR model. The MWA, combined with HSC Deep field survey at $z=6.6$, can detect the signal at large scales if the reionization proceeds relatively slowly (late model), while the SKA has enough sensitivity to detect the signal for all models. Follow-up observations by PFS is very effective to enhance the detectability, especially at small scales. With PFS, the SKA can reach the turnover scale where the sign of cross spectrum changes. However, this may not be true if we use the LAE model in \citet{2018arXiv180100067I} because the enhancement of the cross-spectrum signal at small scales due to larger LAE bias is diminished by one order of magnitude in that case.

To understand the sensitivity curve, we compared error components of cross-correlation measurements including sample variance, thermal noise of radio telescope, shot noise of LAEs and LAE redshift errors. While the sensitivity with the MWA is limited by the thermal noise, sample variance is dominant for the SKA. This indicates that the MWA can improve the S/N ratio by increasing the observing time and/or the number of antennae and that the observing time of 10 hrs is enough for the SKA to detect the signal at large scales. The situation is the same for a case without PFS. Further, we found out that another effective way to increase the S/N ratio is to expand the survey area, rather than to perform deeper observation, if we have an extended LAE survey.

As we mentioned in Sec.4, the turnover scale of the cross-power spectrum is expected to give a typical size of ionized bubbles and can be a important clue to probe the process of reionization. A measurement of the turnover scale would be possible only by the SKA with HSC and PFS as we saw in Fig.~\ref{fig:deep_mid}. We can estimate the accuracy of the determination of the turnover scale by estimating the S/N ratios of individual wavenumber bins varying the bin width. Consequently, for the mid model, the accuracy is estimated to be $\Delta k \sim \pm6\times10^{-3}~{\rm Mpc}^{-1}$. The accuracy apparently depends on EoR model but it seems likely that the turnover scale can be measured by the combination of the SKA, HSC and PFS.

In our analysis, the effects associated with the foreground subtraction were not considered. Although the foreground does not contribute to the average value of cross-correlation measurements, it does contribute to the variance and can degrade the detectability significantly. Thus, a quantitative estimate of this effect is necessary to investigate the feasibility more realistically. This paper aims for investigating the intrinsic feasibility. The effect of foregrounds on the 21cm-LAE cross-correlation is investigated in \citet{2017arXiv170904168Y}.

\section*{Acknowledgement}
We thank Tomoaki Ishiyama for providing us with $N$-body simulation data used in this work. Numerical simulations were carried out on CrayXC30 installed at Center for Computational Astrophysics of National Astronomical Observatory of Japan, NAOJ. The author thanks Cathryn Trott for helpful comments that improved the paper. This work is supported by Grand-in-Aid from the Ministry of Education, Culture, Sports, and Science and Technology (MEXT) of Japan, No.26610048, No.15H05896, No.16H05999, No.17H01110 (KT), No.16J01585(SY), No.17H04827(HY), Bilateral Joint Research Projects of JSPS (KT), a grant from NAOJ, KAKENHI (15H02064) Grant-in-Aid for Scientific Research (A) through Japan Society for the Promotion of Science, and the Centre for All-sky Astrophysics (an Australian Research Council Centre of Excellence funded by grant CE110001020).




%
%


\bsp	
\label{lastpage}
\end{document}